\documentclass[a4paper,11pt]{article}

\usepackage{jheppub} 

\usepackage[utf8]{inputenc}
\usepackage[T1]{fontenc} 
\usepackage{amsfonts}
\usepackage{multirow}
\usepackage{mathrsfs}
\usepackage{graphicx}
\usepackage{amsmath}
\usepackage{amssymb}
\usepackage{adjustbox}
\usepackage{bm}
\usepackage{bbm}
\usepackage{color}
\usepackage{array}
\usepackage{diagbox}
\usepackage{float}
\usepackage{slashed}
\usepackage{url}
\usepackage{ulem}
\usepackage[colorlinks=true,
            linkcolor=blue,
            filecolor=blue,
            anchorcolor=blue,
            urlcolor=blue,
            citecolor=blue
            ]{hyperref}
            
\allowdisplaybreaks[1]

\def\OMIT#1{}

\def\hlinew#1{%
  \noalign{\ifnum0=`}\fi\hrule \@height #1 \futurelet
   \reserved@a\@xhline}
\usepackage{array}
\newcommand{\PreserveBackslash}[1]{\let\temp=\\#1\let\\=\temp}
\newcolumntype{C}[1]{>{\PreserveBackslash\centering}p{#1}}
\newcolumntype{R}[1]{>{\PreserveBackslash\raggedleft}p{#1}}
\newcolumntype{L}[1]{>{\PreserveBackslash\raggedright}p{#1}}

\newcommand{\beq}{\begin{equation}}
\newcommand{\eeq}{\end{equation}}
\newcommand{\bqa}{\begin{eqnarray}}
\newcommand{\eqa}{\end{eqnarray}}

\newcommand\fverb{\setbox\fverbbox=\hbox\bgroup\verb}
\newcommand\fverbdo{\egroup\medskip\noindent%
			\fbox{\unhbox\fverbbox}\ }
\newcommand\fverbit{\egroup\item[\fbox{\unhbox\fverbbox}]}
\newbox\fverbbox

\makeatletter

\newcommand{\Rmnum}[1]{\expandafter\@slowromancap\romannumeral #1@}
\makeatother

\title{\bf \boldmath Rare $W \to B_c + \gamma$ decay up to the NNLO and NLL accuracy in QCD}

\author[a,b,1]{Xin-Qiang Li,\note{Corresponding author.}}
\author[a,c]{Ya-Dong Yang,}
\author[a]{Yu-Dong Zhang}
\author[a]{and Dong-Hui Zheng}

\affiliation[a]{Institute of Particle Physics and Key Laboratory of Quark and Lepton Physics (MOE), Central China Normal University, Wuhan, Hubei 430079, China}
\affiliation[b]{Center for High Energy Physics, Peking University, Beijing 100871, China}
\affiliation[c]{Institute of Particle and Nuclear Physics, Henan Normal University, Xinxiang 453007, China}

\emailAdd{xqli@mail.ccnu.edu.cn}
\emailAdd{yangyd@mail.ccnu.edu.cn}
\emailAdd{ydzhang@mails.ccnu.edu.cn}
\emailAdd{zhengdh@mails.ccnu.edu.cn}

\abstract{We perform a detailed theoretical study of the rare radiative decay of the $W$ boson into a $B_c$ meson and an on-shell photon. The decay amplitude is described by two independent form factors, which are calculated up to the next-to-next-to-leading order (NNLO) in QCD within the nonrelativistic QCD (NRQCD) factorization formalism. Since the two typical energy scales, the $W$-boson mass $m_W$ and the $B_c$-meson mass $m_{B_c}$, involved in the process are widely separated, large logarithms of $m_W^2/m_{B_c}^2$ present in the NRQCD short-distance coefficients are also resummed to all orders in $\alpha_s$ up to the next-to-leading logarithmic (NLL) accuracy, by employing the light-cone factorization approach. Taking into account all these corrections, we then perform a phenomenological exploration of this rare decay. It is found that, relative to the leading-order result, the decay width of the process is reduced by the next-to-leading-order and NNLO corrections, with a net effect of $\sim-19\%$ and of $\sim-31\%$, respectively. Furthermore, the NLL resummation can considerably alter the fixed-order NRQCD predictions, especially for the $\mathcal{O}(\alpha_s)$ correction. We also find that the radiative corrections increase the renormalization scale dependence of the branching fraction, which is however significantly reduced by the NLL resummation. The dependence of the branching fraction on the heavy-quark masses $m_{b,c}$ is also investigated, which shows a monotonic decrease (increase) with $m_c$ ($m_b$).}

\begin{document} 
\maketitle
\flushbottom

\section{Introduction}
\label{sec:introduction}

The $B_c$ meson, being the lightest state composed of a charm quark $c$ and a bottom anti-quark $\bar{b}$ with spin zero, is a unique system that offers distinct challenges and opportunities for both theoretical and experimental studies~\cite{QuarkoniumWorkingGroup:2004kpm,Brambilla:2010cs}. As the $B_c$ meson carries two different heavy flavors, its productions and decays provide a good platform for further understanding the heavy-quark dynamics, which is generally expected to be complementary to those provided by the charmonium and bottonium systems~\cite{Brambilla:2004jw}. The first observation of approximately 20 $B_c$ events in the $B_c \to J/\psi \ell \nu$ decay by the CDF collaboration~\cite{CDF:1998ihx,CDF:1998axz} opens up the possibility of experimental study of the system. Recently, the $B_c$-meson productions have been extensively studied in $pp$~\cite{Chang:1992jb,Berezhnoy:2019yei}, $e^+e^-$~\cite{Zheng:2015ixa,Berezhnoy:2016etd,Zheng:2017xgj,Zheng:2019gnb,Zhang:2021ypo,Zhan:2022etq}, $ep$~\cite{Bi:2016vbt}, and $\gamma\gamma$~\cite{Chen:2020dtu} collisions. The $B_c$ meson is stable against both strong and electromagnetic interactions, and can only decay weakly through the following three distinct mechanisms: a) the $b$-quark decay with the $c$ quark as a spectator, which accounts for about $20\%$ of the total decay width; b) the $c$-quark decay with the $b$ quark as a spectator, which accounts for about $70\%$ of the total decay width; and c) the simultaneous annihilation of both $b$ and $c$ quarks, which accounts for about $10\%$ of the total decay width. All these three mechanisms are interesting and could offer new insights into the weak decay mechanism of heavy flavors~\cite{QuarkoniumWorkingGroup:2004kpm,Brambilla:2010cs}. The current status and future prospects for the $B_c$-meson studies at the Large Hadron Collider (LHC) could be found, e.g., in refs.~\cite{Gouz:2002kk,Gao:2010zzc,HeavyFlavorAveragingGroupHFLAV:2024ctg,ParticleDataGroup:2024cfk}.

From the time of its discovery in 1983~\cite{UA1:1983crd,UA2:1983tsx}, the $W$ boson has been extensively studied in $p\bar{p}$, $pp$ and $e^+e^-$ interactions. Precision knowledge of the $W$-boson properties is of great importance in testing the internal consistency of the Standard Model (SM) and probing new physics beyond it. The $W$ boson decays predominantly into a quark-antiquark pair that manifests as two hadronic jets, with a branching fraction of approximate two-thirds~\cite{ParticleDataGroup:2024cfk}. Although being quite rare, the radiative hadronic decays $W\to M+\gamma$, where the quark-antiquark pair turns instead into a single meson $M$, are also interesting. They provide a significant opportunity to explore the coupling of the $W$ boson with the photon and, more importantly, offer a powerful testing ground for the strongly coupled Quantum Chromodynamics (QCD) regime~\cite{Arnellos:1981gy,Keum:1993eb,Grossman:2015cak,Feng:2019meh,Ishaq:2019zki,Bakos:2022lek,Beneke:2023nmj,Bagdatova:2023etj}. Furthermore, they could be used to measure the $W$-boson mass through fully reconstructed and highly resolved final states~\cite{Jones:2020bvu,Mangano:2014xta}. Experimentally, however, no such kinds of processes have been observed so far, and only upper limits on the branching ratios of some channels like $W \to D_s + \gamma$~\cite{CDF:1998kzn,LHCb:2022kta} and $W \to \pi + \gamma$~\cite{CDF:2011jwr,CMS:2020oqe,ATLAS:2023jfq} exist. On the other hand, a huge number of $W$-boson events expected at the high-luminosity LHC (HL-LHC)~\cite{Azzi:2019yne}, the Future Circular collider (FCC-ee and FCC-hh)~\cite{FCC:2018evy,FCC:2018vvp}, and the Circular Electron Positron Collider (CEPC)~\cite{CEPCStudyGroup:2018ghi} will significantly facilitate the experimental studies of these rare radiative hadronic decays. 

Theoretically, the rare exclusive processes $W \to M + \gamma$ can be calculated up to a given order in perturbative theory in electroweak and/or QCD couplings. First calculations were performed at the leading order (LO)~\cite{Arnellos:1981gy,Keum:1993eb}, but more recent results exist at the next-to-leading-order (NLO) in QCD~\cite{Grossman:2015cak,Feng:2019meh,Ishaq:2019zki}; for a recent review, we refer the readers to ref.~\cite{dEnterria:2023wjq}. In particular, due to the mass hierarchy $m_W \gg m_M$, with $m_W$ the $W$-boson mass and $m_M$ the meson mass, these processes have been investigated in ref.~\cite{Grossman:2015cak} by using the light-cone (LC) factorization formalism~\cite{Lepage:1980fj,Chernyak:1983ej}, according to which the decay amplitude can be factorized into convolutions of the hard-scattering kernels that are calculable in perturbative QCD with the meson light-cone distribution amplitudes (LCDAs) that encode the infrared physics of the final-state hadron formation. This formalism also facilitates the resummation of large logarithms of $m_W^2/\mu_0^2$, with $\mu_0\sim 1~\mathrm{GeV}$ a typical hadronic scale, to all orders in QCD coupling $\alpha_s$~\cite{Lepage:1979zb,Efremov:1978rn,Efremov:1979qk}. The exclusive radiative $Z$-boson decays to an $S$- or a $P$-wave quarkonium have been studied in refs.~\cite{Guberina:1980dc,Wang:2013ywc,Huang:2014cxa,Bodwin:2017pzj,Chung:2019ota,Sang:2022erv,Sang:2023hjl}, finding that the leading (LL) and the next-to-leading logarithmic (NLL) resummation have crucial impacts on the fixed-order theoretical predictions, primarily due to the large logarithms of $m_Z^2/\mu_0^2$ present in the short-distance coefficients (SDCs) of the decay amplitudes.

In this work, we will perform a detailed theoretical study of the $W\to B_c+\gamma$ decay, by combining the LC and the nonrelativistic QCD (NRQCD) factorization~\cite{Brambilla:2004jw,Bodwin:1994jh} approaches. As the $B_c$ meson is a heavy meson with quarkonium-like properties that is composed of two different species of heavy quarks, it is natural to employ the NRQCD factorization formalism to tackle this exclusive $B_c$-meson production process. On the other hand, since $m_W$ is much larger than the $B_c$-meson mass $m_{B_c}$, the LC factorization (also known as the collinear factorization)~\cite{Lepage:1980fj,Chernyak:1983ej} is also a viable approach. In ref.~\cite{Feng:2019meh}, the same process has been studied up to the NLO in $\alpha_s$ by combining the LC and NRQCD factorization approaches. We will extend their calculations to the next-to-next-to-leading order (NNLO) in $\alpha_s$, but still at the lowest order in the typical velocity $v$ of the $c/\bar{b}$ quark inside the $B_c$-meson state, within the NRQCD factorization formalism. To address the issue of large logarithms of $m_W^2/m_{B_c}^2$ present in the NRQCD SDCs, we will employ the LC factorization formalism to refactorize the NRQCD SDCs and utilize the celebrated Efremov-Radyushkin-Brodsky-Lepage (ERBL) evolution equation~\cite{Lepage:1979zb,Efremov:1978rn,Efremov:1979qk} to resum these large logarithms. Concretely, we will perform both the LL and the NLL resummation for the leading-twist helicity amplitude, i.e., resumming both the $\alpha_s^n\ln^n(m_W^2/m_{B_c}^2)$ and $\alpha_s^{n+1}\ln^n(m_W^2/m_{B_c}^2)$ terms to all orders in $\alpha_s$. Taking into account all these corrections, we will then perform a phenomenological exploration of the $W\to B_c+\gamma$ decay. It is found that, relative to the LO result, the decay width of the process is reduced by the $\mathcal{O}(\alpha_s)$ and $\mathcal{O}(\alpha_s^2)$ corrections, with a net effect of $\sim-19\%$ and of $\sim-31\%$ respectively. Furthermore, the NLL resummation can considerably alter the fixed-order NRQCD predictions, especially for the $\mathcal{O}(\alpha_s)$ correction. We also find that the radiative corrections increase the renormalization scale dependence of the branching fraction, which is however significantly reduced by the NLL resummation. The variations of the branching fraction with respect to the heavy-quark masses $m_{b,c}$ are also investigated, which shows a monotonic decrease (increase) with $m_c$ ($m_b$).

The rest of this paper is organized as follows. In section~\ref{sec-formula}, after setting up our notation, we present the general Lorentz decomposition of the decay amplitude and the unpolarized decay rate of $W\to B_c+\gamma$ expressed in terms of the two nonvanishing form factors. In section~\ref{sec-NRQCD}, we employ the NRQCD factorization formalism to factorize the helicity amplitudes, introduce the procedure and techniques to compute the SDCs, and present the SDCs at various perturbative orders. Section~\ref{sec-LL} is devoted to the LC refactorization for the NRQCD SDCs. Furthermore, the LL and NLL resummations are formulated and explicitly carried out. Since the analytical expressions are rather cumbersome, only the formal expressions are presented in this section, with the final analytical results for each formula provided in appendix~\ref{appendix-convolutions}. Detailed phenomenological analysis is then presented in section~\ref{sec-phen}. Finally, we summarize in section~\ref{sec-summary}. For convenience, the explicit expressions of the ERBL kernels are given in appendix~\ref{appendix-BL-kernels}. In appendix~\ref{appendix-branching-ratios-mc-mb}, the resulting decay widths and branching fractions at various levels of accuracy are tabulated for different charm- and bottom-quark pole masses.

\section{Amplitude decomposition and decay width}
\label{sec-formula}

Throughout this paper, we work in the rest frame of the $W$ boson, and assign the momenta of the outgoing $B_c$ and photon by $p$ and $k$, respectively. The law of momentum conservation dictates that the four-momentum of the $W$ boson is determined by $q=p+k$. They are subject to the on-shell conditions, with $q^2=m_W^2$, $p^2=m_{B_c}^2$, and $k^2=0$. The polarization vector of the $W$ boson is denoted by $\varepsilon_{W}(S_z)$, with $S_z$ the spin-$z$ component, while the photon polarization vector by $\varepsilon_\gamma(\lambda)$, with $\lambda=\pm1$ the photon helicity. In accordance with Lorentz invariance and taking into account the fact that an on-shell photon must be transversely polarized, the decay amplitude of $W\to B_c+\gamma$ can be decomposed as~\cite{Grossman:2015cak,Feng:2019meh}
\bqa \label{eq-helicity-selection-rule}
\mathcal{A}_\lambda(W^+ \to B_c^+ + \gamma) = \mathcal{F}_1\,\varepsilon_W(S_z) \cdot \varepsilon_\gamma^*(\lambda) + \frac{\mathcal{F}_2}{m_W^2}\, i \epsilon_{\mu \nu \alpha \beta} \varepsilon_W^{\mu} \left(S_z\right) \varepsilon_\gamma^{* \nu}(\lambda) p^\alpha k^\beta,
\eqa
where $\mathcal{F}_1$ and $\mathcal{F}_2$ are the two scalar form factors that encode all the nontrivial QCD dynamics involved in the process. Our main task is then to compute these two form factors from the first principles of QCD, as will be detailed in Sections~\ref{sec-NRQCD} and \ref{sec-LL}. For example, they can be expressed in terms of overlap integrals of calculable hard-scattering kernels with the $B_c$-meson LCDA in the LC factorization approach. Here we use the convention $\epsilon^{0123}=-\epsilon_{0123}=1$ for the totally antisymmetric Levi-Civita tensor $\epsilon_{\mu \nu \alpha \beta}$. Since parity invariance is violated by weak interaction, the two helicity amplitudes of the process are actually independent of each other, and they can be expressed via the linear combination of these two form factors.

Squaring eq.~\eqref{eq-helicity-selection-rule} and averaging (summing) over the polarization states of the $W$ boson (photon), we can write the unpolarized decay width of $W^+ \to B_c^+ + \gamma$ in the $W$-bsoson rest frame as~\footnote{Throughout this paper, we will focus on the process $W^+\to B_c^+ + \gamma$. Since $B_c^+$ is composed of a $c\bar{b}$ pair while $B_c^-$ is a $\bar{c}b$ bound state, all the results presented here can be transplanted to the case of $W^-\to B_c^- + \gamma$ by making the substitutions $e_c\leftrightarrow -e_b$, $m_c\leftrightarrow m_b$, and $x_0\leftrightarrow\bar{x}_0$.}
\bqa\label{eq-gen-rate-helicity}
\Gamma(W^+ \to B_c^+ + \gamma) &=& \frac{1}{2J+1}\frac{1}{2 m_W} \frac{1}{8 \pi}\frac{2|\mathbf{p}|}{m_W} \sum_{\lambda=\pm1}|\mathcal{A}_\lambda(W^+ \to B_c^+ + \gamma)|^2\nonumber \\
&=& \frac{|\mathbf{p}|}{24 \pi m_W^2}\left(\left|\mathcal{A}_{+1}\right|^2+\left|\mathcal{A}_{-1}\right|^2\right).
\eqa
Here $J=1$ denotes the spin of the $W$ boson, and $|\mathbf{p}|$ is the magnitude of the three-momentum of the $B_c$ meson in the $W$-boson rest frame, which is readily determined by
\beq
|\mathbf{p}| = \frac{\lambda^{1/2}(m_W^2,m_{B_c}^2,0)}{2m_W}=\frac{m_W^2-m_{B_c}^2}{2m_W},
\eeq
with the K\"{a}ll\'{e}n function defined by $\lambda(x,y,z)=x^2+y^2+z^2-2xy-2yz-2zx$. Note that, due to the absence of parity invariance in weak interaction, there is no straightforward symmetry linking the two helicity amplitudes $\mathcal{A}_{\pm1}$. Equivalently, we can express the unpolarized decay width in terms of the two scalar form factors defined in eq.~\eqref{eq-helicity-selection-rule} as
\beq
\Gamma(W^+ \to B_c^+ + \gamma) = \frac{|\mathbf{p}|}{12 \pi m_W^2}\left(\left|\mathcal{F}_1\right|^2+\frac{|\mathbf{p}|^2}{m_W^2}\left|\mathcal{F}_2\right|^2\right).
\eeq

\section{Form factors in the NRQCD factorization formalism}
\label{sec-NRQCD}

For the rare $W \to B_c + \gamma$ process, the $B_c$ meson can be treated as a heavy meson with quarkonium-like properties, which means that the charm quark and the bottom antiquark have to be created in a very short distance. This guarantees that the asymptotic freedom of QCD can be safely invoked to compute the quark-level hard-scattering amplitude in perturbation theory. On the other hand, to have a substantial probability to form a $B_c$ bound state, the relative motion between the charm quark and the bottom antiquark must be necessarily slow. This implies that the quark-level amplitude should be insensitive to the relative momentum between the heavy quark and antiquark, and can be therefore expanded as a power series of the heavy-quark relative velocity $v$. All the binding dynamics of the process is embedded into a multiplicative long-distance factor, which is usually modelled by the $B_c$-meson wave function at the origin. Such a physical picture lies at the heart of the NRQCD factorization~\cite{Brambilla:2004jw,Bodwin:1994jh}, and will be adopted here.

\subsection{NRQCD factorization for the helicity amplitudes}

According to the NRQCD factorization formalism~\cite{Brambilla:2004jw,Bodwin:1994jh}, the helicity amplitudes $\mathcal{A}_{\lambda}$ of $W^+ \to B_c^+ + \gamma$ can be factorized at the lowest order in $v$ as~\cite{Feng:2019meh}
\begin{equation}\label{eq-nrqcd}
\mathcal{A}_\lambda(W^+ \to B_c^+ + \gamma) = \sqrt{2 m_{B_c}}\frac{\langle\mathcal{O}_{B_c}\rangle}{\sqrt{2 N_c}\, 2 m_{\rm g.m.}} \tilde{\mathcal{A}}_\lambda(W^+ \to c\bar{b}({}^1 S_0^{(1)}) + \gamma),
\end{equation}
where $N_c=3$ is the number of colors. The nonrelativistically normalized long-distance matrix element (LDME) $\langle\mathcal{O}_{B_c}\rangle$ is defined by
\begin{equation} \label{eq:LDME_def}
\langle\mathcal{O}_{B_c}\rangle=\bigl|\langle B_c|\psi_c^{\dagger} \chi_b|0\rangle\bigr|,
\end{equation}
where $\psi_c^{\dagger}$ and $\chi_b$ are the Pauli spinor fields creating a charm quark and a bottom antiquark in NRQCD, respectively. The quark-level helicity amplitudes $\tilde{\mathcal{A}}_\lambda$ on the right-hand side of eq.~(\ref{eq-nrqcd}) can be obtained by replacing the physical $B_c$ state with the free $c\bar{b}$ state carrying the same ${}^1 S_0^{(1)}$ quantum numbers as that of the leading Fock component of the $B_c$ meson. In eq.~(\ref{eq-nrqcd}), we have also introduced the auxiliary variable $m_{\rm g.m.} \equiv \sqrt{m_b m_c}$, which represents the geometric mean of the charm-quark mass $m_c$ and the bottom-quark mass $m_b$. The factor $\sqrt{2m_{B_c}}$ appears on the right-hand side of eq.~(\ref{eq-nrqcd}), because we have adopted the relativistic normalization for the $B_c$ state, but use the conventional nonrelativistic normalization for the LDME $\langle\mathcal{O}_{B_c}\rangle$ defined by eq.~\eqref{eq:LDME_def}. In this work, we will not consider the relativistic correction, and it is therefore reasonable to take the approximation $m_{B_c}\approx m_b+m_c$.  

To facilitate our later discussions, let us detach some irrelevant common factors from the two form factors $\mathcal{F}_{1,2}$, and introduce the following two dimensionless short-distance coefficients (SDCs):
\begin{equation} \label{eq-helicity-amplitude-1}
\mathcal{C}_{1,2}=\frac{\sin\theta_W}{4\pi \alpha V_{cb} f_{B_c}^{(0)}}\,\mathcal{F}_{1,2},
\end{equation}
where $V_{cb}$ is the Cabibbo–Kobayashi–Maskawa (CKM) matrix element involved in the process, $\theta_{W}$ the weak mixing angle, and $\alpha=e^2/(4\pi)$ the fine structure constant. The LO decay constant $f_{B_c}^{(0)}$ of the $B_c$ meson is defined by
\begin{equation}
f_{B_c}^{(0)}=\sqrt{\frac{2}{m_{B_c}}}\,\langle\mathcal{O}_{B_c}\rangle.
\end{equation}
It is convenient to parametrize the dimensionless SDCs in powers of $\alpha_s$ as
\begin{equation} \label{eq-sdcs-expand}
\mathcal{C}_i = \mathcal{C}_i^{(0)} + \frac{\alpha_s}{\pi} \mathcal{C}_{i}^{(1)} + \left(\frac{\alpha_s}{\pi}\right)^2 \left[\frac{\beta_0}{4}\ln\frac{\mu_R^2}{m_{\rm g.m.}^2}\,\mathcal{C}_{i}^{(1)} + \frac{\gamma_{B_c}}{2} \ln\frac{\mu_\Lambda^2}{m_{\rm g.m.}^2}\,\mathcal{C}_i^{(0)} + \mathcal{C}_{i}^{(2)}\right] + \mathcal{O}(\alpha_s^3),
\end{equation}
where $\mu_R$ and $\mu_\Lambda$ represent the renormalization and the factorization scale, respectively. $\beta_0 = (11/3) C_A - (4/3) T_F n_f$ is the one-loop coefficient of the QCD beta function, where $C_A=3$, $T_F=1/2$, and $n_f$ is the number of active quark flavors. The explicit $\ln\mu_R^2$ term in eq.~\eqref{eq-sdcs-expand} is deduced from the renormalization group invariance of the physical amplitude. $\gamma_{B_c}$ represents the anomalous dimension associated with the NRQCD bilinear current $\psi_c^{\dagger} \chi_b$ carrying the same quantum number ${}^1S_0$ as that of the $B_c$ meson, and its explicit expression in the modified minimal subtraction ($\overline{\rm MS}$) scheme reads~\cite{Chen:2015csa,Tao:2023mtw}
\begin{equation}\label{eq-ano-dim}
\gamma_{B_c}=-\pi^2\left[\frac{a^2+6a+1}{2(a+1)^2} C_F^2+\frac{C_F C_A}{2}\right],
\end{equation}
where $a=m_c/m_b$, and $C_F=(N_c^2-1)/(2N_c)=4/3$ is the quadratic Casimir invariant of the QCD gauge group $SU(N_c)$. The $\mu_\Lambda$ dependence of the SDCs in eq.~\eqref{eq-sdcs-expand} should be exactly cancelled by that of the NRQCD LDME, as required by the NRQCD factorization.

We now describe briefly our procedure to evaluate the quark-level helicity amplitudes $\tilde{\mathcal{A}_\lambda}$, from which the SDCs $\mathcal{C}_{1,2}$ can be straightforwardly extracted. At the LO in the heavy-quark relative velocity $v$, we have simply $m_{B_c}\approx m_b+m_c$, and the momenta of the comoving charm quark and bottom antiquark inside the $B_c$ state can be assigned, respectively, as
\begin{equation}
p_c=\frac{m_c}{m_{B_c}}p \approx \frac{m_c}{m_b+m_c}p=x_0 p, \qquad p_b=\frac{m_b}{m_{B_c}}p \approx \frac{m_b}{m_b+m_c}p=\bar{x}_0 p,
\end{equation}
where $x_0=\frac{m_c}{m_b+m_c}$, and $\bar{x}_0=1-x_0$. The on-shell condition enforces that $p_c^2=m_c^2$, $p_b^2=m_b^2$, and
\begin{equation}
p^2 = m_{B_c}^2 \approx (m_b+m_c)^2, \qquad p \cdot k \approx \frac{m_W^2-(m_b+m_c)^2}{2}.
\end{equation}
During our calculation, we first evaluate the helicity amplitudes $\tilde{\mathcal{A}_\lambda}$ of the quark-level process $W^+\to c\bar{b} + \gamma$, and then employ the covariant spin and color projector technique to extract the spin- and color-singlet component of the $c\bar{b}$ pair. Explicitly, we replace the product of two quark spinors by the spin- and color-singlet projector~\cite{Bodwin:2002cfe}:
\begin{equation} \label{eq:Bc_projector}
v(p_b) \bar{u}(p_c) \rightarrow \bar{\Pi}_0=\frac{1}{\sqrt{8 m_c m_b}}\bigl(\slashed{p}_b-m_b\bigr) \gamma^5\bigl(\slashed{p}_c+m_c\bigr) \otimes \frac{\boldsymbol{1}}{\sqrt{N_c}},
\end{equation}
where the two spinors are normalized relativistically. In this way, the quark-level helicity amplitudes of $W^+ \to c\bar{b}({}^1 S_0^{(1)}) + \gamma$ relevant for the $W^+ \to B_c^+ + \gamma$ decay can be readily projected out with the aid of
\begin{eqnarray}
&& \tilde{\mathcal{A}}_\lambda(W^+ \to c\bar{b} + \gamma) 
= \varepsilon_{W,\mu}\, \varepsilon_{\gamma,\nu}^{*} \, \bar{u}(p_c) \tilde{\mathcal{A}}_\lambda^{\mu \nu} v(p_b) 
= \varepsilon_{W,\mu}\, \varepsilon_{\gamma,\nu}^{*} \operatorname{Tr}\left[v(p_b) \bar{u}(p_c) \tilde{\mathcal{A}}_\lambda^{\mu \nu}\right] \nonumber \\[0.2cm]
&\rightarrow& \tilde{\mathcal{A}}_\lambda(W^+ \to c\bar{b}({}^1 S_0^{(1)}) + \gamma) = \varepsilon_{W,\mu}\, \varepsilon_{\gamma,\nu}^{*} \operatorname{Tr}\left[\bar{\Pi}_0 \tilde{\mathcal{A}}_\lambda^{\mu \nu}\right],
\end{eqnarray}
where $\tilde{\mathcal{A}}_\lambda^{\mu \nu}$ represent the quark-level amplitudes of $W^+\to c\bar{b} + \gamma$ with the external quark spinors stripped off.

\subsection{SDCs through \texorpdfstring{$\mathcal{O}(\alpha_s^2)$}{O(alphas2)}}

\begin{figure}[t]
    \centering
    \includegraphics[scale=0.45]{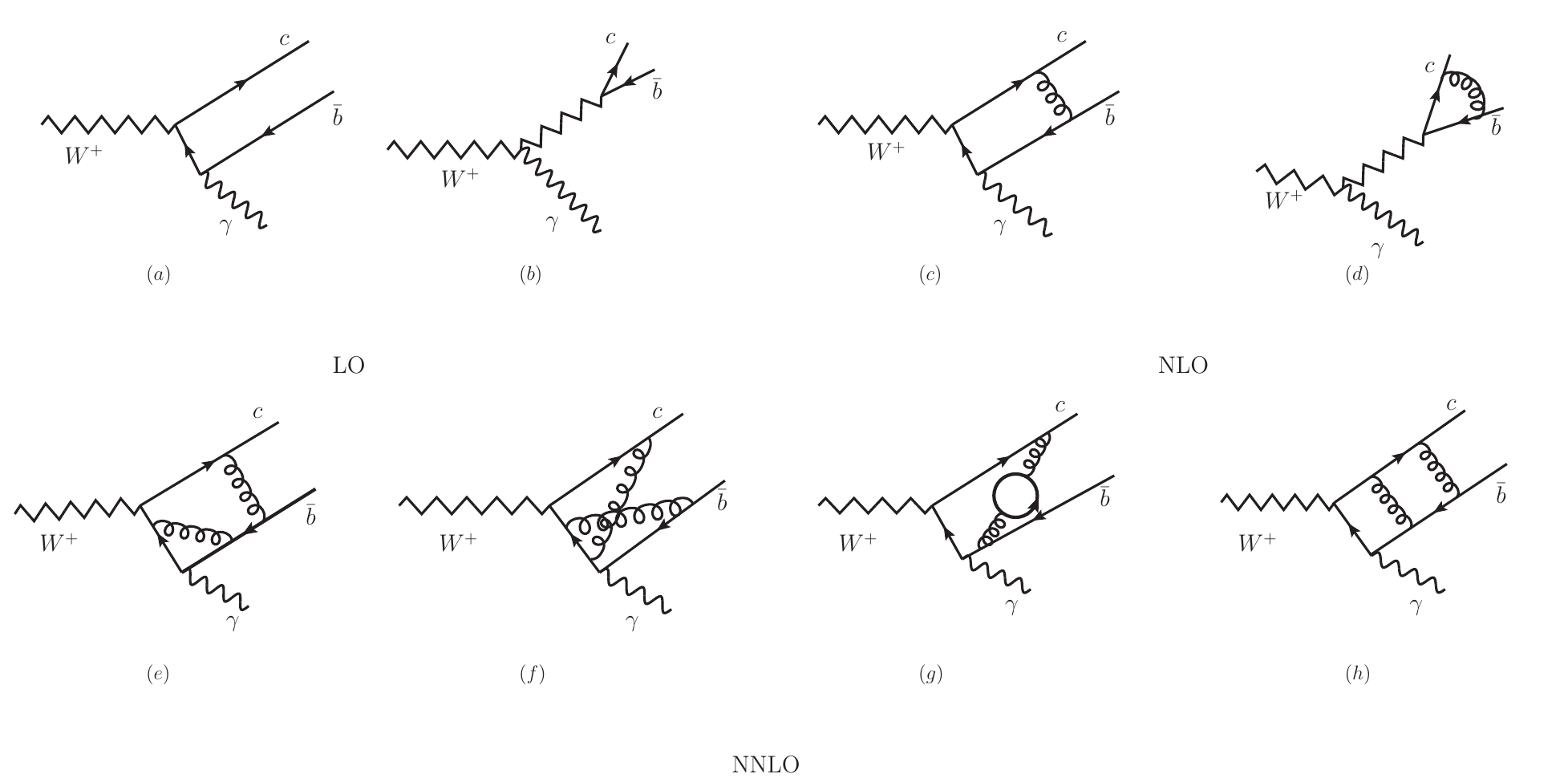}
    \caption{Some representative quark-level Feynman diagrams for the process $W^+\to B_c^+ + \gamma$ up to $\mathcal{O}(\alpha_s^2)$. \label{fig-feynman-diagram}}
\end{figure}

We proceed to calculate the SDCs $\mathcal{C}_{1,2}$ up to $\mathcal{O}(\alpha_s^2)$ within the NRQCD factorization formalism. The quark-level Feynman diagrams and the corresponding amplitudes are generated using the package \texttt{FeynArts}~\cite{Kublbeck:1990xc,Hahn:2000kx}. We show in figure~\ref{fig-feynman-diagram} some representative quark-level Feynman diagrams for the process $W^+\to B_c^+ + \gamma$ up to $\mathcal{O}(\alpha_s^2)$. Employing the spin and color projector defined by eq.~\eqref{eq:Bc_projector} to enforce the $c\bar{b}$ pair in ${}^1 S_0^{(1)}$ for the $B_c$ meson, we can then obtain the hadron-level amplitudes order by order in $\alpha_s$ with the aid of the packages \texttt{FeynCalc}~\cite{Mertig:1990an,Shtabovenko:2016sxi,Shtabovenko:2020gxv,Shtabovenko:2023idz} and \texttt{FormLink}~\cite{Feng:2012tk}. The LO results of the reduced SDCs introduced in eq.~\eqref{eq-helicity-amplitude-1} can be straightforwardly obtained from the first two LO Feynman diagrams in figure~\ref{fig-feynman-diagram}, together with the one obtained from figure~\ref{fig-feynman-diagram}($a$) by attaching the emitted photon to the charm quark rather to the bottom antiquark line. Explicitly, we have
\begin{eqnarray}
\mathcal{C}_1^{(0)} & = & \frac{1}{4 \sqrt{2}}\left(\frac{e_c}{x_0}-\frac{e_b}{\bar{x}_0}\right)-\frac{1}{2 \sqrt{2}\left(1-r^2\right)}, 
\label{sdcs-lo-1} \\[0.2cm]
\mathcal{C}_2^{(0)} & =& \frac{1}{2 \sqrt{2}\left(r^2-1\right)}\left(\frac{e_c}{x_0}+\frac{e_b}{\bar{x}_0}\right), \label{sdcs-lo-2}
\end{eqnarray}
where $r=\frac{m_{B_c}}{m_W}$, and $e_c=+\frac{2}{3}$ and $e_b=-\frac{1}{3}$ are the electric charges (in unit of the elementary charge $e$) of the charm and bottom quarks, respectively.

Once going beyond the LO in $\alpha_s$, we have to perform a detailed matching calculation from QCD onto NRQCD to extract the SDCs. This can also be achieved by directly computing the hard region of the loop integrals in full QCD in the context of strategy of regions~\cite{Beneke:1997zp,Smirnov:2002pj}, because contributions from all the other regions in full QCD always have a one-to-one correspondence to those in NRQCD and are therefore cancelled out during the matching~\cite{Beneke:1997jm}. Here we will adopt such a shortcut to extract the SDCs. To this end, we need to evaluate 9 one-loop and 186 two-loop Feynman diagrams, with some representative ones shown already in figure~\ref{fig-feynman-diagram}. All the calculations are conducted in the Feynman-'t Hooft gauge, and the dimensional regularization scheme is employed to regularize both ultraviolet (UV) and infrared (IR) divergences. After implementing the on-shell renormalization scheme for the heavy-quark masses and fields~\cite{Broadhurst:1991fy,Melnikov:2000zc,Fael:2020bgs}, and the $\overline{\mathrm{MS}}$ renormalization scheme for the QCD coupling, the UV poles are exactly cancelled out, while an uncancelled single IR pole still remains. Such a symptom is a common feature of the NRQCD factorization, which has been encountered many times in the NNLO perturbative calculations involving heavy quarkonia (see, e.g., refs.~\cite{Chen:2015csa,Tao:2023mtw,Beneke:1997jm}). This IR pole can be factored into the NRQCD LDMEs, so that the final NRQCD SDCs become IR finite. We have verified numerically with high precision that the coefficient of the remaining IR pole equals to one-quarter of the anomalous dimension in eq.~(\ref{eq-ano-dim}), as required by the NRQCD factorization. 

As the dimensional regularization scheme is adopted throughout this work, we have to fix the prescription of $\gamma_5$ in $D\neq 4$ dimensions. It is well-known that, in this scheme, the anticommutation relation, $\{\gamma^\mu,\gamma_5\}=0$, and the cyclicity of Dirac traces, $\operatorname{Tr}(\gamma^\mu\cdots\gamma^\nu)=\operatorname{Tr}(\gamma^\nu\gamma^\mu\cdots)$, cannot be satisfied simultaneously. In practical computation, the naive-$\gamma_5$ scheme~\cite{Kreimer:1989ke,Korner:1991sx}, which keeps the anticommutation relation of $\gamma_5$ in $D\neq 4$ dimensions and avoids the spurious anomaly, is frequently adopted. However, because of the lack of the cyclicity of Dirac traces, one must fix the reading point for a fermion loop with an odd number of $\gamma_5$, and all the Dirac traces must be read starting at the same reading point~\cite{Korner:1991sx}. In this work, we will select the vertex of the $B_c$ meson as the reading point. For a similar treatment, we refer the interesting readers to refs.~\cite{Sang:2022erv,Sang:2023hjl}.

For the one- and two-loop integrals involved, we will utilize the packages \texttt{Apart}~\cite{Feng:2012iq} and \texttt{FIRE}~\cite{Smirnov:2008iw,Smirnov:2014hma,Smirnov:2019qkx,Smirnov:2023yhb} to reduce them into linear combinations of some independent master integrals (MIs). In this way, we end up with 10 one-loop MIs, which are computed analytically using \texttt{Package-X}~\cite{Patel:2015tea,Patel:2016fam}, and roughly 380 two-loop MIs, whose evaluation is a rather challenging task. Fortunately, a new powerful algorithm, dubbed ``Auxiliary Mass Flow'', has recently been proposed by Liu and Ma~\cite{Liu:2017jxz,Liu:2020kpc,Liu:2021wks}. Its main idea is to compute the MIs by setting up and solving a set of differential equations with respect to an auxiliary mass variable $\eta$ introduced into certain propagators, with the boundary conditions at $\eta\to \infty$ given simply by vacuum bubble integrals that can be more easily calculated by using other methods~\cite{Davydychev:1992mt,Broadhurst:1993mw,Tarasov:1997kx,Martin:2003qz,Schroder:2005va,Baikov:2010hf,Lee:2011jt} or by the method itself. Therefore, zero input of boundary conditions can be realized with the iterative applications of this method. Remarkably, these differential equations can be solved iteratively with very high numerical precision. In this work, we will utilize the newly released package \texttt{AMFlow}~\cite{Liu:2022chg,Liu:2022mfb} to compute all the two-loop MIs numerically. 

The analytical expressions of the NLO coefficients $\mathcal{C}_i^{(1)}$ in eq.~\eqref{eq-sdcs-expand} can be readily obtained. Here, instead of presenting their complete cumbersome expressions, we present merely their asymptotic expansions in $r\to 0$:
\begin{eqnarray} 
\mathcal{C}_1^{(1)} & = & \left\{-\frac{C_F}{16 \sqrt{2}} \frac{e_c}{x_0}\left[\left(2 \ln x_0+3\right)\ln \left(-r^2+i \epsilon\right)+\ln ^2 x_0+3 \left(\frac{x_0}{\bar{x}_0}+5\right) \ln x_0+3 \ln \bar{x}_0\right.\right. \nonumber\\[0.15cm]
&& \left.-4 \mathrm{Li}_2(x_0)+\frac{2 \pi^2}{3}+9\right] \left.-\binom{e_c \rightarrow e_b}{x_0 \rightarrow \bar{x}_0}\right\}-\frac{3 C_F}{8 \sqrt{2}}\left[\left(x_0-\bar{x}_0\right) \ln \frac{x_0}{\bar{x}_0}-2\right], \label{sdcs-nlo-1} \\[0.3cm]
\mathcal{C}_2^{(1)} & = & \frac{C_F}{8 \sqrt{2}} \frac{e_c}{x_0}\bigg[\left(2 \ln x_0+3\right)\ln \left(-r^2+i \epsilon\right)+\ln ^2 x_0+\left(\frac{x_0}{\bar{x}_0}+5\right) \ln x_0+3 \ln \bar{x}_0+\frac{2 \pi^2}{3} \nonumber\\[0.15cm]
&& -4 \mathrm{Li}_2(x_0)+9\bigg]+\binom{e_c \rightarrow e_b}{x_0 \rightarrow \bar{x}_0}, \label{sdcs-nlo-2}
\end{eqnarray}
where $\mathrm{Li}_2(x)$ is the dilogarithm function defined by $\mathrm{Li}_2(x)=-\int_0^x \text{ln}(1-t)/t\,dt$, and $\epsilon$ is an infinitesimal positive real number. It is worth noting that our results of $\mathcal{C}_{2}^{(0)}$, $\mathcal{C}_{1}^{(1)}$, and $\mathcal{C}_{2}^{(1)}$ in eqs.~(\ref{sdcs-lo-1})--(\ref{sdcs-nlo-2}) are completely consistent with $f_{2}^{(0)}$, $f_{1}^{(1)}$, and  $f_2^{(1)}$ in eqs.~(11) and (15) of ref.~\cite{Feng:2019meh}, but $\mathcal{C}_{1}^{(0)}$ differs from $f_1^{(0)}$ by a factor of $1/(1-r^2)$ in the second term.

It is rather challenging for us to deduce the analytical expressions of the two-loop coefficients $\mathcal{C}_i^{(2)}$ in eq.~\eqref{eq-sdcs-expand}. Therefore, we turn instead to present their high-precision numerical values. To this end, we take as inputs $m_W=80.3692$~GeV from the latest particle data group (PDG)~\cite{ParticleDataGroup:2024cfk}, and the charm- and bottom-quark pole masses $m_c = 1.69$~GeV and $m_b=4.80$~GeV, which are converted from the $\overline{\rm MS}$ running masses $\bar{m}_c(\bar{m}_c)=1.28$ GeV and $\bar{m}_b(\bar{m}_b)=4.18$ GeV~\cite{ParticleDataGroup:2024cfk} at the two-loop level in QCD with the aid of the package \texttt{RunDec}~\cite{Chetyrkin:2000yt,Herren:2017osy}. We tabulate in table~\ref{tab-sdcs-mc-1.69} the numerical values of the SDCs 
$\mathcal{C}_1^{(n)}$ and $\mathcal{C}_2^{(n)}$ for $n=0,1,2$, respectively. For the sake of reference, the explicit dependence of the two SDCs $\mathcal{C}_{1,2}^{(2)}$ on $n_l$, $n_c$ and $n_b$ are kept, where $n_l$ denotes the numbers of the light-quark flavors, while $n_c=1$ and $n_b=1$ signify the number of the charm and of the bottom quark respectively. The variations of our theoretical results with respect to the charm- and bottom-quark pole masses will be investigated in appendix~\ref{appendix-branching-ratios-mc-mb}.

\begin{table}[t]
    \renewcommand{\arraystretch}{1.3}
    \tabcolsep=0.46cm
    \centering
    \begin{tabular}{|c|c|c|c|}
    \hline\hline
    & $n=0$ & $n=1$ & $n=2$\\
    \hline
    \multirow{2}*{$\mathcal{C}_1^{(n)}$} 
    & 
    \multirow{2}*{$0.176$}&\multirow{2}*{$-0.108-0.327i$}&$-(5.647-3.072i)-(0.374-0.193i)n_l$\\
    &&& $-(0.125-0.189i)n_c-(0.130-0.167i)n_b$\\
    \hline
    \multirow{2}*{$\mathcal{C}_2^{(n)}$}
    & \multirow{2}*{$-0.751$}&\multirow{2}*{$2.372-0.119i$}&$38.672+4.553i+(0.152-0.704i)n_l$\\
    &&&$-(0.433+0.699i)n_c+(0.143-0.660i)n_b$\\
    \hline\hline
    \end{tabular}
    \caption{Numerical values of the dimensionless SDCs $\mathcal{C}_1^{(n)}$ and $\mathcal{C}_2^{(n)}$ evaluated at different orders of $\alpha_s$ in the NRQCD factorization formalism. For the sake of reference, we have kept explicitly the $n_l$, $n_c$ and $n_b$ dependences of the two SDCs $\mathcal{C}_{1,2}^{(2)}$, where $n_l$ denotes the number of the light quarks, while $n_c=1$ and $n_b=1$ the numbers of charm and bottom quarks respectively. \label{tab-sdcs-mc-1.69}} 
\end{table}

\section{LC factorization at leading twist for the SDCs}
\label{sec-LL}

As mentioned in section~\ref{sec:introduction}, the rare process $W\to B_c + \gamma$ can also be analyzed in the LC factorization approach~\cite{Lepage:1980fj,Chernyak:1983ej}. This is based on the observation that, for an asymptotically large $m_W$, the outgoing $B_c$ meson moves nearly with the speed of light, and both the quark and antiquark inside it are dictated by light-like kinematics. This implies that the quark-level hard-scattering amplitude is sensitive neither to the quark masses $m_{b,c}$ nor the transverse momentum $p_\perp$ carried by the quark and antiquark. Thus, the decay amplitude of $W\to B_c + \gamma$ can be expanded in powers of $m_{b,c}^2/m_W^2$ and $p_\perp^2/m_W^2$, with all the hadron structure effects lumped into the $B_c$-meson LCDAs. Explicitly, we can factorize the decay amplitude into convolution integrals of the perturbatively calculable hard-scattering kernels with the leading-twist LCDA of the $B_c$ meson order by order in $\alpha_s$, at the leading power in an expansion of $\Lambda_\mathrm{QCD}/m_W$, where $\Lambda_\mathrm{QCD}$ denotes the typical strong-interaction scale. The same formalism also facilitates the resummation of large logarithms of $m_W^2/\mu_0^2$ to all orders in perturbation theory, with the aid of the ERBL evolution equation of the $B_c$-meson LCDA~\cite{Lepage:1979zb,Efremov:1978rn,Efremov:1979qk}. In this section, we will demonstrate that the NRQCD SDCs can be refactorized in the LC factorization formalism, and the NLL resummation has a considerable impact on the fixed-order NRQCD predictions. 

\subsection{LC factorization}
\label{subsec-LL}

Following the spirit of refs.~\cite{Ma:2006hc,Bell:2008er,Jia:2008ep,Jia:2010fw,Wang:2013ywc,Xu:2016dgp}, we can refactorize the NRQCD SDCs in the LC factorization approach as
\begin{equation} \label{eq:SDC_in_LC}
\mathcal{C}_i = \int_0^1 d x\, T_i(x, m_W,\mu)\, \hat{\phi}(x,m_{B_c}, \mu)+\mathcal{O}(m_{B_c}^2/m_W^2)\,,
\end{equation}
where $x$ denotes the LC momentum fraction carried by the charm quark in the $B_c$ state, and $\mu$ is the factorization scale. The hard-scattering kernels $T_i$ and the leading-twist LCDA $\hat\phi$ are perturbatively calculable around the scales $m_W$ and $m_{B_c}$, respectively. Up to $\mathcal{O}(\alpha_s)$, we have the formal expansions:
\begin{eqnarray} 
    T_i(x, m_W,\mu) &=& T_i^{(0)}(x)+\frac{\alpha_s(\mu)}{4 \pi} T_i^{(1)}(x,m_W,\mu)+\mathcal{O}(\alpha_s^2)\,, \label{eq:T_expansion} \\[0.2cm]
    \hat{\phi}(x,m_{B_c},\mu) &=& \hat{\phi}^{(0)}(x)+\frac{\alpha_s(\mu)}{4 \pi} \hat{\phi}^{(1)}(x,m_{B_c},\mu)+\mathcal{O}(\alpha_s^2)\,, \label{eq:phi_expansion}
\end{eqnarray}
where the explicit expressions of the coefficients of the hard-scattering kernels are given, respectively, as~\cite{Feng:2019meh}
\begin{subequations}\label{T-kernel}
    \begin{eqnarray}
    T_1^{(0)}(x) & = & \frac{1}{4\sqrt{2}}\biggl[\left(\frac{e_c}{x}-\frac{e_b}{\bar{x}}\right)-2\biggr]\,, \\[0.2cm]
    T_2^{(0)}(x) & = & -\frac{1}{2\sqrt{2}}\left(\frac{e_c}{x}+\frac{e_b}{\bar{x}}\right)\,, \\[0.2cm]
    T_1^{(1)}(x,m_W,\mu) & =& \frac{C_F}{4\sqrt{2}} \Biggl\{\frac{e_c}{x}\left[(2 \ln x+3) \left(\ln \frac{m_W^2}{\mu^2}-i\pi\right) + \ln ^2 x+(8\Delta-3)\frac{x}{\bar{x}} \ln x-9\right]\nonumber\\[0.15cm]
    && -\left(\begin{array}{c}e_c\rightarrow e_b\\x \leftrightarrow \bar{x}\end{array}\right)+8\Delta\Biggr\}\,,\\[0.2cm]
    T_2^{(1)}(x,m_W, \mu) & =& -\frac{C_F}{2\sqrt{2}} \Bigg\{\frac{e_c}{x}\left[(2 \ln x+3) \left(\ln \frac{m_W^2}{\mu^2}-i\pi\right) + \ln ^2 x+(8\Delta-1)\frac{x}{\bar{x}} \ln x-9\right]\nonumber\\[0.15cm]
    && +\left(\begin{array}{c}e_c\rightarrow e_b\\x \leftrightarrow \bar{x}\end{array}\right)\Bigg\}\,.
    \end{eqnarray}
\end{subequations}
Here $\bar{x}=1-x$, and $\Delta=0$ in the naive dimensional regularization (NDR) scheme with anticommuting $\gamma_5$, while $\Delta=1$ in the 't Hooft-Veltman (HV) scheme~\cite{tHooft:1972tcz}, in which $\gamma_5=i\gamma^0\gamma^1\gamma^2\gamma^3$ anticommutes with the four matrices $\gamma^\mu$ with $\mu\in \{0, 1, 2, 3\}$, and commutes with the remaining $D-4$ Dirac matrices. Note that, once the same prescription adopted in the hard-scattering kernels are followed when computing the $B_c$-meson LCDA, such a scheme dependence should be exactly cancelled out by that of the leading-twist LCDA of the $B_c$ meson, as can be checked by comparing eqs.~\eqref{T-kernel} with \eqref{phi-kernel}.

At the leading power in the heavy-quark relative velocity $v$ and up to $\mathcal{O}(\alpha_s)$, the coefficients of the leading-twist LCDA of the $B_c$ meson are given, respectively, as~\cite{Bell:2008er,Jia:2008ep,Jia:2010fw,Xu:2016dgp}
\begin{subequations}\label{phi-kernel}    
    \begin{eqnarray}
    \hat{\phi}^{(0)}\left(x\right) & =& \delta(x-x_0)\,, \\[0.3cm]
    \hat{\phi}^{(1)}\left(x, m_{B_c},\mu\right) & = & 2C_F\Biggl\{\biggl(\ln\frac{\mu^2}{m_{B_c}^2\left(x_0-x\right)^2}-1\biggr) \Biggl[\frac{x_0+\bar{x}}{x_0-x} \frac{x}{x_0} \theta(x_0-x) +\left(\begin{array}{c}x \leftrightarrow \bar{x} \\ x_0 \leftrightarrow \bar{x}_0\end{array}\right)\Biggr]\Biggr\}_{+} \nonumber\\[0.15cm]
    && +C_F\Biggl\{\biggl(\frac{4x \bar{x}}{\left(x_0-x\right)^2}\biggr)_{++} + 8 \Delta \left[\frac{x}{x_0} \theta(x_0-x)+\left(\begin{array}{c}x \leftrightarrow \bar{x} \\ x_0 \leftrightarrow \bar{x}_0\end{array}\right)\right]_{+}\nonumber\\[0.15cm]
    && +2 \delta^{\prime}(x-x_0) \left(2 x_0 \bar{x}_0 \ln \frac{x_0}{\bar{x}_0}
    +x_0-\bar{x}_0\right)\Biggr\}\nonumber\\[0.15cm]
    && +C_F \Biggl[4 \Delta+\left(-6+3\left(x_0-\bar{x}_0\right) \ln \frac{x_0}{\bar{x}_0}\right)\Biggr]\delta(x-x_0).
    \end{eqnarray}
\end{subequations}
Note that the terms proportional to $\delta(x-x_0)$ in the NLO coefficient $\hat\phi^{(1)}$ actually contribute to the one-loop correction to the $B_c$-meson decay constant $f_{B_c}$~\cite{Xu:2016dgp}. It is also interesting to note that the momentum fractions carried by the charm quark and the bottom antiquark can be reshuffled to a highly asymmetric configuration through exchanging energetic collinear gluons, making the NLO coefficient $\hat\phi^{(1)}$ have a nonvanishing support in the full range from 0 to 1 and develop a long tail. This is in sharp contrast to the infinitely narrow LO coefficient $\hat\phi^{(0)}$, which reflects simply that the partition of the $B_c$ momentum by the charm quark and the bottom antiquark is commensurate with their mass ratios, as expected at the leading-power expansion in NRQCD, in which the charm quark and the bottom antiquark can be assumed to be at rest relative to each other in the $B_c$ rest frame.

Finally, let us remind that the ``$+$'' and ``$++$'' prescriptions in the NLO coefficient $\hat\phi^{(1)}$ of eq.~(\ref{phi-kernel}) should be understood in the sense of distributions. Specifically, for a test function $f(x)$ that has a smooth behavior near $x=x_0$, its convolutions with the ``$+$'' and ``$++$'' functions are defined, respectively, by~\cite{Bell:2008er,Jia:2008ep,Jia:2010fw,Xu:2016dgp}
\begin{equation}
\begin{aligned}
\int_0^1 dx\, [g(x)]_{+} f(x) & = \int_0^1 dx\, g(x)\left[f(x)-f(x_0)\right],\\[0.3cm]
\int_0^1 dx\, [g(x)]_{++} f(x) & = \int_0^1 dx\, g(x)\left[f(x)-f(x_0)-f^{\prime}(x_0)\left(x-x_0\right)\right].
\end{aligned}
\end{equation}

\subsection{NLL resummation with the ERBL kernels}
\label{subsec-resum}

The leading-twist LCDA $\hat{\phi}(x,m_{B_c},\mu)$ of the $B_c$ meson obeys the celebrated ERBL evolution equation~\cite{Lepage:1979zb,Efremov:1978rn,Efremov:1979qk}
\begin{equation} \label{eq:ERBLeq}
\mu^2 \frac{d}{d \mu^2} \hat\phi(x, m_{B_c},\mu) = \int_0^1 d y\, V(x, y; \alpha_s(\mu))\, \hat\phi(y,m_{B_c}, \mu)\,,
\end{equation}
where the perturbatively calculable evolution kernel can be expanded in $\alpha_s$ as
\begin{eqnarray} \label{eq:ERBL-kernel}
V(x, y; \alpha_s(\mu)) = \frac{\alpha_s(\mu)}{4 \pi} V^{(0)}(x, y) + \left(\frac{\alpha_s(\mu)}{4 \pi}\right)^2 V^{(1)}(x, y) + \mathcal{O}(\alpha_s^3)\,.
\end{eqnarray}
The explicit expressions of the one- and two-loop coefficients are collected in appendix~\ref{appendix-BL-kernels}.

The formal solution of the differential-integral evolution equation given by eq.~\eqref{eq:ERBLeq} can be written as
\begin{eqnarray} \label{eq:evolution_LCDA}
\hat\phi(x,m_{B_c},\mu) &=& U(\mu,\mu_0)\ast\hat\phi(x,m_{B_c},\mu_0)\,,
\end{eqnarray}
where the symbol ``$\ast$'' indicates an appropriate convolution over the LC momentum fraction, and the evolution function is defined by
\begin{equation}
U(\mu,\mu_0) = \mathrm{P}\exp\left\{\int_{\alpha_s(\mu_0)}^{\alpha_s(\mu)}d\alpha_s\frac{V(x, y; \alpha_s)}{\beta(\alpha_s)}\right\}\,,
\end{equation}
with $\mathrm{P}$ standing for the path ordering in $\alpha_s$, and $\beta(\alpha_s)$ being the QCD beta function. Explicitly, the evolution function can be expanded as a Dyson series as
\begin{eqnarray}
U(\mu,\mu_0) \nonumber &=& 1+\int_{\alpha_s(\mu_0)}^{\alpha_s(\mu)}d\alpha_s\frac{V(x, y; \alpha_s)}{\beta(\alpha_s)} +\int_{\alpha_s(\mu_0)}^{\alpha_s(\mu)}d\alpha_{s1}\frac{V(x, y; \alpha_{s1})}{\beta(\alpha_{s1})}\ast\int_{\alpha_s(\mu_0)}^{\alpha_{s1}}d\alpha_{s2}\frac{V(x, y; \alpha_{s2})}{\beta(\alpha_{s2})}\nonumber\\[0.15cm] 
&& + \cdots 
+\int_{\alpha_s(\mu_0)}^{\alpha_s(\mu)}d\alpha_{s1}\frac{V(x, y; \alpha_{s1})}{\beta(\alpha_{s1})}\ast\int_{\alpha_s(\mu_0)}^{\alpha_{s1}}d\alpha_{s2}\frac{V(x, y; \alpha_{s2})}{\beta(\alpha_{s2})}\ast\cdots\nonumber\\[0.15cm] 
&& \hspace{1.0cm} \ast\int_{\alpha_s(\mu_0)}^{\alpha_{s(n-1)}}d\alpha_{sn}\frac{V(x, y; \alpha_{sn})}{\beta(\alpha_{sn})}+\cdots\,.
\end{eqnarray}

Plugging eq.~\eqref{eq:evolution_LCDA} into eq.~\eqref{eq:SDC_in_LC}, we can then write the renormalization-group improved SDCs for the $W\to B_c+\gamma$ decay at the leading power in the LC factorization formalism as
\begin{equation}
\mathcal{C}_i = \int_0^1 d x\, T_i\left(x, m_W, m_W\right) U\left(m_W, m_{B_c}\right) \hat\phi\left(x, m_{B_c}, m_{B_c}\right)+\mathcal{O}\left(m_{B_c}^2/m_W^2\right)\,.
\end{equation}

\subsubsection{Truncated LL and NLL resumed SDCs}

Let us truncate the perturbative expansion of the evolution function $U(\mu,\mu_0)$ up to $\mathcal{O}(\alpha_s^2)$, and obtain 
\begin{eqnarray} \label{eq:expandedNLL}
U(\mu,\mu_0) &\approx& 1+\frac{\alpha_s(\mu)}{4\pi}\ln\frac{\mu^2}{\mu_0^2}V^{(0)}(x,y) + \left(\frac{\alpha_s(\mu)}{4\pi}\right)^2 \Biggl[\ln\frac{\mu^2}{\mu_0^2} V^{(1)}(x,y) \nonumber \\[0.15cm] 
&& \hspace{1.1cm} + \frac{1}{2}\ln^2\frac{\mu^2}{\mu_0^2} \left(\int_0^1 d z V^{(0)}(x,z) V^{(0)}(z,y) + \beta_0 V^{(0)}(x,y)\right) \Biggr]\,.
\end{eqnarray}
Here we have used the renormalization group evolution of the QCD coupling $\alpha_s(\mu)$ accurate up to the NLL level,
\begin{equation} \label{eq:as-run}
\alpha_s(\mu) = \alpha_s(\mu_0)\left[1+\beta_0\frac{\alpha_s(\mu_0)}{4\pi}\ln\frac{\mu^2}{\mu_0^2}+\beta_1\left(\frac{\alpha_s(\mu_0)}{4\pi}\right)^2 \ln\frac{\mu^2}{\mu_0^2}\right]^{-1}\,,
\end{equation}
and its perturbative expansion up to $\mathcal{O}(\alpha_s^2)$,
\begin{equation}\label{eq:as-run-expand}
\alpha_s(\mu) = \alpha_s(\mu_0)\left[1-\frac{\alpha_s(\mu_0)}{4\pi} \Biggl(\beta_0\ln\frac{\mu^2}{\mu_0^2}\right) +\left(\frac{\alpha_s(\mu_0)}{4\pi}\right)^2 \left(\beta_0^2\ln^2\frac{\mu^2}{\mu_0^2}-\beta_1\ln\frac{\mu^2}{\mu_0^2}\right)+\mathcal{O}(\alpha_s^3)\Biggr]\,,
\end{equation}
where $\beta_0$ and $\beta_1$ are the first two coefficients in a perturbative expansion of the QCD beta function in $\alpha_s$, $\beta(\alpha_s)=-\alpha_s\sum_{n=0}^{\infty}\beta_n\left(\frac{\alpha_s}{4\pi}\right)^{n+1}$, with $\beta_0$ given already below eq.~\eqref{eq-sdcs-expand} and $\beta_1=(34/3)C_A^2-(20/3)C_A T_F\,n_f -4 C_F T_F\,n_f$.

Equipped with eqs.~\eqref{eq:T_expansion}, \eqref{eq:phi_expansion} and \eqref{eq:expandedNLL}, we can write the LL and NLL resumed SDCs truncated up to $\mathcal{O}(\alpha_s^2)$, respectively, as 
\begin{subequations} \label{eq:NLL-expand}
\begin{eqnarray} 
\widetilde{\mathcal{C}}_i^{\rm LL} &\approx& \int_0^1 d x\, T_i^{(0)}(x) \phi^{(0)}(x)+\frac{\alpha_s\left(m_W\right)}{4 \pi} \ln \frac{m_W^2}{m_{B_c}^2} \int_0^1 d x d y\, T_i^{(0)}(x) V^{(0)}(x, y) \phi^{(0)}(y)  \nonumber \\[0.15cm]
&& +\frac{1}{2}\left(\frac{\alpha_s\left(m_W\right)}{4 \pi}\right)^2 \ln ^2 \frac{m_W^2}{m_{B_c}^2}\Biggl[\int_0^1 d x d y d z\, T_i^{(0)}(x) V^{(0)}(x, y) V^{(0)}(y, z) \phi^{(0)}(z) \nonumber\\[0.15cm]
&& \hspace{4.2cm} +\beta_0 \int_0^1 d x d y\, T_i^{(0)}(x) V^{(0)}(x, y) \phi^{(0)}(y)\Biggr]\,, \\[0.3cm]
\widetilde{\mathcal{C}}_i^{\rm NLL} &\approx& \widetilde{\mathcal{C}}_i^{\rm LL}\left(m_W, m_{B_c}\right)+\frac{\alpha_s(m_W)}{4 \pi}\int_0^1 d x\, T_i^{(1)}(x)\phi^{(0)}(x) +\frac{\alpha_s(m_{B_c})}{4 \pi}\int_0^1 d x\, T_i^{(0)}(x) \phi^{(1)}(x) \nonumber \\[0.15cm]
&& +\left(\frac{\alpha_s\left(m_W\right)}{4 \pi}\right)^2 \ln \frac{m_W^2}{m_{B_c}^2}\Biggl[\int_0^1 d x d y \left(T_i^{(0)}(x) V^{(0)}(x, y) \phi^{(1)}\left(y\right)\right.  \nonumber \\[0.15cm]
&& \hspace{2.7cm} \left. +T_i^{(1)}(x) V^{(0)}(x, y) \phi^{(0)}(y)+T_i^{(0)}(x) V^{(1)}(x, y) \phi^{(0)}(y)\right)\Biggr]\,.
\end{eqnarray}
\end{subequations}
Since the resulting analytical expression of each term in eq.~(\ref{eq:NLL-expand}) is rather cumbersome, we present in this subsection only the formal expressions as given above, with the full analytical result of each term listed in appendix~\ref{appendix-convolutions}. Note that the dependence of the hard-scattering kernels $T_i$, the leading-twist LCDA $\hat \phi$, and the evolution kernel $V$ with respect to the $\gamma_5$ scheme cancels with each other eventually in the above expansion of ${\widetilde{\mathcal{C}}}_i^{\rm NLL}$, as expected. 

\subsubsection{Resummation to all orders in \texorpdfstring{$\alpha_s$}{alphas} with Gegenbauer polynomial expansion}

The LL resummation to all orders in $\alpha_s$ is commonly done with the aid of the Gegenbauer polynomials by noting that such polynomials are the eigenfuctions of the LO evolution kernel $V^{(0)}(x,y)$. To perform the NLL resummation to all orders in $\alpha_s$, on the other hand, we have to take into account both the diagonal and non-diagonal parts of the evolution kernel $U_{n,k}$ in the Gegenbauer moment space, which will be introduced in eq.~\eqref{eq:Unk_matrix}.

The Gegenbauer expansion of the leading-twist LCDA of the $B_c$ meson is given by
\begin{equation}
\hat\phi(x, m_{B_c},\mu) \equiv \sum_{n=0}^{\infty} \hat\phi_{n}(\mu) x(1-x) C_n^{(3 / 2)}(2 x-1)\,,	
\end{equation}
where $C^{(3/2)}_n(x)$ are the Gegenbauer polynomials of rank $3/2$, and the Gegenbauer moments are defined by
\begin{equation}
\hat\phi_{n}(\mu)=\frac{4(2 n+3)}{(n+1)(n+2)}\int_0^1 d x\, C_n^{(3 / 2)}(2 x-1) \, \hat\phi(x, m_{B_c}, \mu) \,.	
\end{equation}
Note that $\hat\phi_{n}(\mu)$ have a similar perturbative expansion in $\alpha_s$ around the scale $m_{B_c}$ as that of $\hat\phi(x,m_{B_c},\mu)$ (cf. eq.~\eqref{eq:phi_expansion}). Solving the ERBL evolution equation in the Gegenbauer moment space, we have formally~\cite{Jia:2008ep}
\begin{equation} \label{eq:Unk_matrix}
\hat\phi_{n}(\mu)=\sum_{k=0}^n U_{n, k}\left(\mu, \mu_0\right) \hat\phi_{k}(\mu_0)\,,
\end{equation}
where the matrix elements of the evolution kernel $U_{n,k}$ in the Gegenbauer moment space up to $\mathcal{O}(\alpha_s)$ can be found, e.g., in ref.~\cite{Agaev:2010aq}, the convention of which will be followed here. 

Similarly, we have the Gegenbauer expansion of the hard-scattering kernels
\begin{equation}
T_i(x, m_W, \mu) = \sum_{n=0}^{\infty} \frac{4(2 n+3)}{(n+1)(n+2)}\, T_{i,n}(\mu)\, C_n^{(3 / 2)}(2 x-1)\,,
\end{equation}
with
\begin{equation}
T_{i,n}(\mu)=\int_0^1 d x\, x(1-x) C_n^{(3 / 2)}(2 x-1)\, T_i(x, m_W,\mu)\,.	
\end{equation}
It is also noted that $T_{i,n}(\mu)$ have a similar perturbative expansion in $\alpha_s$ around the scale $m_W$ as that of $T_i(x,m_W,\mu)$ (cf. eq.~\eqref{eq:T_expansion}). Thus, we can write the LL and NLL resumed SDCs to all orders in $\alpha_s$, respectively, as
\begin{subequations}
\begin{eqnarray} 
\mathcal{C}_i^{\rm LL} &=& \sum_{n=0}^{\infty}\sum_{k=0}^\infty T_{i,n}^{(0)}(m_W)\, U_{n,k}^\mathrm{LO}(m_W,m_{B_c})\,\hat\phi_{k}^{(0)}(m_{B_c}) \label{sdc-sum-LL} \\[0.3cm]
\mathcal{C}_i^{\rm NLL} &=& \sum_{n=0}^{\infty}\sum_{k=0}^\infty T_{i,n}(m_W)\, U_{n,k}^\mathrm{NLO}(m_W,m_{B_c})\,\hat\phi_{k}(m_{B_c}) \nonumber \\[0.15cm]
&=& \mathcal{C}_i^{(0,0)}+\frac{\alpha_s(m_W)}{4 \pi} \mathcal{C}_i^{(1,0)} + \frac{\alpha_s\left(m_{B_c}\right)}{4 \pi} \mathcal{C}_i^{(0,1)} \,, \label{sdc-sum-NLL}
\end{eqnarray}
\end{subequations}
where
\begin{equation}\label{eq:NLL-resummation}
\mathcal{C}_i^{(u, v)}=\sum_{n=0}^{\infty}\sum_{k=0}^\infty  T_{i,n}^{(u)}(m_W)\, 
U_{n, k}^\mathrm{NLO}(m_W,m_{B_c})\, \hat\phi_{k}^{(v)}(m_{B_c})\,,
\end{equation}
and the explicit expressions of $U_{n,k}^\mathrm{(N)LO}$ could be found in ref.~\cite{Agaev:2010aq}. It is worth emphasizing again that the resulting SDCs $\mathcal{C}_i^{\rm NLL}$ are indeed independent of the $\gamma_5$ scheme used during the calculations of the hard-scattering kernels $T_i$, the leading-twist LCDA $\hat \phi$, as well as the evolution kernel $V^{(1)}$.

\subsection{SDCs by combining the NRQCD prediction and the LC resummation}

We proceed to compute the SDCs by combining the NRQCD prediction and the LC resummation. To avoid any double counting, we should subtract the $\alpha_s^n \ln^n r^2$ terms from the fixed-order NRQCD predictions for the LL resummation, and both the $\alpha_s^n \ln^n r^2$ and $\alpha_s^{n+1} \ln^{n} r^2$ terms for the NLL resummation. To this end, it is convenient to introduce the following notations:
\begin{subequations}
	\bqa \label{eq-NNLO-LL}
	\mathcal{C}_{i}^{{\rm LO+LL}}&=&\mathcal{C}_{i}^{\rm LO}-
	{\widetilde{\mathcal{C}}}_{i}^{{\rm LL}}\big|_{\alpha_s^0}+\mathcal{C}_{i}^{ \rm LL},\\[0.2cm]
	\mathcal{C}_{i}^{{\rm NLO+(N)LL}}&=&\mathcal{C}_{i}^{\rm NLO}-
	{\widetilde{\mathcal{C}}}_{i}^{{\rm (N)LL}}\big|_{\alpha_s^1}+\mathcal{C}_{i}^{\rm (N)LL},\\[0.2cm]
	\mathcal{C}_{i}^{{\rm NNLO+(N)LL}}&=&\mathcal{C}_{i}^{\rm NNLO}-
	{\widetilde{\mathcal{C}}}_{i}^{{\rm (N)LL}}\big|_{\alpha_s^2}+\mathcal{C}_{i}^{\rm (N)LL},
	\eqa
\end{subequations}
where the superscripts `LO', `NLO' and `NNLO' in the NRQCD SDCs $\mathcal{C}_{i}$ (cf. eq.~\eqref{eq-sdcs-expand}) represent the results accurate up to $\mathcal{O}(\alpha_s^0)$, $\mathcal{O}(\alpha_s^1)$ and $\mathcal{O}(\alpha_s^2)$, respectively. $\mathcal{C}_{i}^{\rm (N)LL}$ denote the (N)LL resumed SDCs to all orders in $\alpha_s$ (cf. eqs.~\eqref{sdc-sum-LL} and (\ref{sdc-sum-NLL}), while ${\widetilde{\mathcal{C}}}_{i}^{{\rm (N)LL}}\big|_{\alpha_s^n}$ indicate the truncated (N)LL resumed SDCs ${\widetilde{\mathcal{C}}}_{i}^{{\rm (N)LL}}$ (cf. eq.~\eqref{eq:NLL-expand}) up to $\mathcal{O}(\alpha_s^n)$.

\begin{figure}[t]
	\centering
	\includegraphics[scale=0.88]{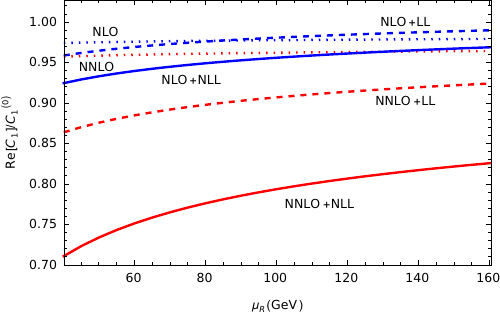}\hfill
	\includegraphics[scale=0.88]{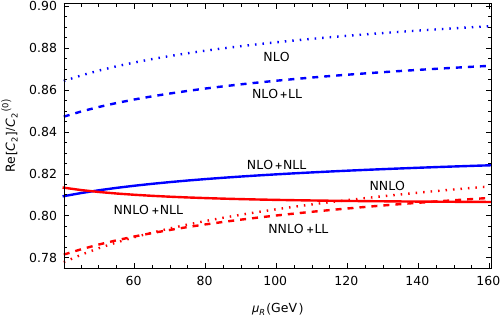}
	\caption{The renormalization scale $\mu_R$ dependence of $\mathrm{Re}\bigl[\mathcal{C}_1\bigr]/\mathcal{C}_1^{(0)}$ and $\mathrm{Re}\bigl[\mathcal{C}_2\bigr]/\mathcal{C}_2^{(0)}$ at various levels of accuracy. \label{fig-sdcs-mu}}
\end{figure}

In figure~\ref{fig-sdcs-mu}, we show the renormalization scale $\mu_R$ dependence of the real parts of the SDCs normalized to the corresponding LO results, $\mathrm{Re}\bigl[\mathcal{C}_1\bigr]/\mathcal{C}_1^{(0)}$ and $\mathrm{Re}\bigl[\mathcal{C}_2\bigr]/\mathcal{C}_2^{(0)}$, at various levels of accuracy. It is observed that the higher-order corrections can significantly alter the LO predictions, but fail to reduce the $\mu_R$ dependence. Interestingly, however, we find that the LL resummation has little impact on the fixed-order predictions for $\mathrm{Re}\bigl[\mathcal{C}_1^{\rm NLO}\bigr]/\mathcal{C}_1^{(0)}$ and $\mathrm{Re}\bigl[\mathcal{C}_2^{\rm NNLO}\bigr]/\mathcal{C}_2^{(0)}$, while the NLL resummation can significantly change the fixed-order results, especially for reducing the $\mu_R$ dependence of $\mathrm{Re}\bigl[\mathcal{C}_2^{\rm NNLO+NLL}\bigr]/\mathcal{C}_2^{(0)}$. In addition, we present in table~\ref{tab-SDCs-squared} our theoretical predictions of the squared SDCs $|\mathcal{C}_1|^2$, $|\mathcal{C}_2|^2$, and $|\mathcal{C}_{\rm tot}|^2$ at various levels of accuracy, where $|\mathcal{C}_{\rm tot}|^2=|\mathcal{C}_1|^2+ \frac{|\mathbf{p}|^2}{m_W^2}|\mathcal{C}_2|^2$. It is worth noting that the sum in eq.~\eqref{eq:NLL-resummation} is formally divergent and, in order to accelerate the convergence, we have used the so-called Abel-Pad\'e method proposed in ref.~\cite{Bodwin:2016edd} to sum the series in eqs.~\eqref{sdc-sum-LL} and \eqref{sdc-sum-NLL}, with a $50\times50$ Pad\'e approximant. 

\begin{table}[t]
    \renewcommand{\arraystretch}{1.3}
    \tabcolsep=0.11cm
    \centering
	\begin{tabular}{|c|c|c|c|c|c|c|c|c|c|}
			\hline \hline
			&LO
			&LO+LL
			&NLO
			&NLO+LL
			&NLO+NLL
			&NNLO
			&NNLO+LL
			&NNLO+NLL\\
			\hline
			$|\mathcal{C}_1|^2$
			&3.11
			&3.90
			&2.98
			&2.98
			&2.80
			&2.89
			&2.52
			&1.89\\
			\hline
			$|\mathcal{C}_2|^2$
			&56.36
			&53.37
			&43.53
			&41.75
			&37.68
			&35.84
			&35.70
			&36.83 \\
			\hline
			$|\mathcal{C}_{\rm tot}|^2$
			&17.03
			&17.08
			&13.73
			&13.29
			&12.11
			&11.74
			&11.34
			&10.98\\
			\hline \hline
	\end{tabular}
    \caption{The squared SDCs~(in unit of $10^{-2}$) at various levels of accuracy, with $m_c=1.69$~GeV, $m_b=4.80$~GeV, and $m_W=80.3692$~GeV. We have also taken $\mu_R=m_W$ and $\mu_\Lambda=1$~GeV. \label{tab-SDCs-squared}}
\end{table}

From table~\ref{tab-SDCs-squared}, we can see that the LL resummation can significantly change the LO NRQCD predictions of $|\mathcal{C}_1|^2$ and $|\mathcal{C}_2|^2$, but has only a negligible impact on $|\mathcal{C}_{\rm tot}|^2$, due to the delicate cancellation between its constructive and destructive corrections to $|\mathcal{C}_1|^2$ and $|\mathcal{C}_2|^2$, respectively. On the other hand, the LL resummation alters only slightly the NLO and NNLO predictions. This is explained by noting that some dominant $\alpha_s^n\ln^n r^2$ contributions have already been included in the NRQCD SDCs calculated at the higher orders in $\alpha_s$; specifically, the $\alpha_s\ln r^2$ and $\alpha^2_s\ln^2 r^2$ contributions have been included in the $\mathcal{O}(\alpha_s)$ and $\mathcal{O}(\alpha_s^2)$ NRQCD SDCs, respectively. However, unlike the case of the LL resummation, we find that the NLL resummation is always significant for the fixed-order NRQCD predictions at various orders in $\alpha_s$.

\section{Phenomenology}
\label{sec-phen}

For our phenomenological analysis, we take as inputs $\sin^2\theta_W=0.231$, $m_W=80.3692$~GeV, $m_{B_c}=6.2749$~GeV, $|V_{cb}|=(41.1\pm1.2)\times10^{-3}$ and fix the fine structure constant at $\alpha(m_W)=1/128$, all being taken from the latest PDG~\cite{ParticleDataGroup:2024cfk}. The QCD coupling $\alpha_s$ at different renormalization scales is evaluated with the aid of the package \texttt{RunDec}~\cite{Chetyrkin:2000yt,Herren:2017osy}. The default value of the renormalization scale is chosen at $\mu_R=m_W$, at which $\alpha_s(m_W) = 0.1205$, and we have varied $\mu_R$ from $m_W/2$ to $2 m_W$ to estimate the theoretical uncertainties in computing the higher-order perturbative QCD corrections. In addition, we approximate the NRQCD LDME at $\mu_\Lambda=1$~GeV by the Schr\"{o}dinger radial wave function at the origin,
\begin{equation} \label{eq-ldme-wave-funciton}
	|\langle \mathcal{O}_{B_c} \rangle|^2\approx \frac{N_c}{2\pi} \left|R_{1S,c\bar{b}}(0)\right|^2
	=\frac{N_c}{2\pi}\times 1.642~\mathrm{GeV}^{3},
\end{equation}
where the $1S$ radial wave function at the origin, evaluated in the Buchm\"uller-Tye potential model, is taken from ref.~\cite{Eichten:1995ch}. Finally, the total decay width of the $W$ boson is taken as $\Gamma_W=2.085\pm0.042$~GeV~\cite{ParticleDataGroup:2024cfk}.

Taking the heavy-quark pole masses $m_c=1.69$ GeV and $m_b=4.80$ GeV, we tabulate in table~\ref{tab:decay-width-ratio} the unpolarized decay widths and branching fractions of $W\to B_c+\gamma$ at various levels of accuracy. Here we have included the uncertainties from the CKM matrix element $|V_{cb}|$ and the total width $\Gamma_{W}$, as well as the variation of the renormalization scale within the range $\mu_R \in [m_W/2, 2 m_W]$. It should be emphasized that the input value of the Schr\"odinger wave function may largely affect the theoretical predictions of these observables. As an illustration, when $\left|R_{1S,c\bar{b}}(0)\right|^2$ varies from $1.642$ to $3.184~\rm GeV^3$~\cite{Eichten:1995ch}, the resulting central values of the decay rates may be changed by roughly a factor of two. It is interesting to note that, relative to the LO result, both the $\mathcal{O}(\alpha_s)$ and $\mathcal{O}(\alpha_s^2)$ corrections to the decay width are negative, with a net effect of $\sim-19\%$ and of $\sim-31\%$, respectively. Compared to the LL resummation, the NLL resummation reduces significantly the fixed-order NRQCD theoretical predictions. This is different from that observed in the rare radiative $Z$-boson decays~\cite{Sang:2022erv,Sang:2023hjl}, which finds that the LL resummation alters significantly the fixed-order theoretical predictions, whereas the NLL resummation changes only slightly the results. 

\begin{table}[t]
    \renewcommand{\arraystretch}{1.3}
    \tabcolsep=1.0cm
    \centering
	\begin{tabular}{|c|c|c|}
		\hline \hline
		Accuracy level
            & $\Gamma~[\,\rm eV\,]$
		& ${\rm Br}~[\,\times 10^{-10}\,]$ \\
		\hline
		LO
		&$0.492_{-0.028}^{+0.029}$
		&$2.359_{-0.133-0.047}^{+0.143+0.049}$ \\
		\hline
		LO+LL
		&$0.493_{-0.028}^{+0.029}$
		&$2.367_{-0.133-0.047}^{+0.143+0.049}$\\
		\hline
		NLO
		&$0.397_{-0.022-0.011}^{+0.024+0.009}$
		&$1.902_{-0.104-0.037-0.051}^{+0.118+0.040+0.041}$\\
		\hline
		NLO+LL
		&$0.384_{-0.021-0.012}^{+0.023+0.010}$
		&$1.841_{-0.101-0.035-0.058}^{+0.114+0.039+0.047}$\\
		\hline
		NLO+NLL
		&$0.350_{-0.020-0.009}^{+0.021+0.008}$
		&$1.678_{-0.092-0.032-0.045}^{+0.104+0.035+0.037}$\\
		\hline
		NNLO
		&$0.339_{-0.019-0.013}^{+0.021+0.011}$
		&$1.627_{-0.088-0.031-0.063}^{+0.102+0.035+0.054}$\\
		\hline
		NNLO+LL
		&$0.328_{-0.018-0.015}^{+0.020+0.012}$
		&$1.571_{-0.085-0.030-0.079}^{+0.099+0.034+0.059}$\\
		\hline
		NNLO+NLL
		&$0.317_{-0.018-0.005}^{+0.019+0.006}$
		&$1.522_{-0.084-0.030-0.026}^{+0.094+0.032+0.028}$\\
		\hline \hline
	\end{tabular}
    \caption{The unpolarized decay widths and branching fractions of $W\to B_c+\gamma$ at various levels of accuracy. The second column represents the decay widths, with the first and second errors arising from the CKM matrix element $|V_{cb}|$ and the renormalization scale $\mu_R$, respectively. The third column corresponds to the branching fractions, where the first, second, and third errors originate from $|V_{cb}|$, $\Gamma_W$, and $\mu_R$, respectively. \label{tab:decay-width-ratio}}
\end{table}

\begin{figure}[t]
    \centering
    \vspace{0.5cm}
    \includegraphics[scale=0.98]{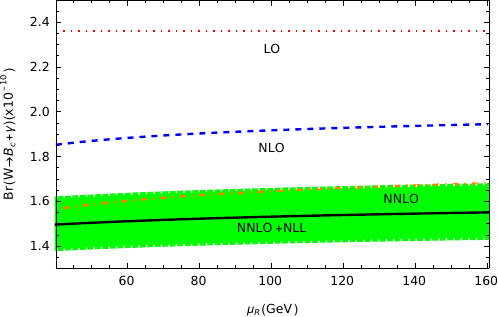}
    \caption{The branching fraction of $W\to B_c+\gamma$ as a function of the renormalization scale $\mu_R$. The green band denotes the uncertainty from $\Gamma_W$ and $|V_{cb}|$. \label{fig-branching-ratio}}
\end{figure}

In order to comprehensively investigate the renormalization scale dependence of our predictions, we have further plotted in figure~\ref{fig-branching-ratio} the branching ratio of $W\to B_c+\gamma$ as a function of the renormalization scale $\mu_R$, where the green band corresponds to the uncertainty associated with the total decay width $\Gamma_W$ and the CKM matrix element $|V_{cb}|$. It is observed that, due to the large scale set by the $W$-boson mass, the $\mu_R$ dependence of the branching ratio is not particularly obvious. Furthermore, by combining the third error in the third column of table~\ref{tab:decay-width-ratio}, it remains evident that the NNLO theoretical predictions do not reduce the $\mu_R$ dependence of the NLO results. In contrast, upon taking into account the NLL resummation, we find that the $\mu_R$ dependence of the branching ratio can be effectively reduced, as can be seen from figure~\ref{fig-branching-ratio}.

\begin{figure}[t]
    \centering
    \includegraphics[scale=0.88]{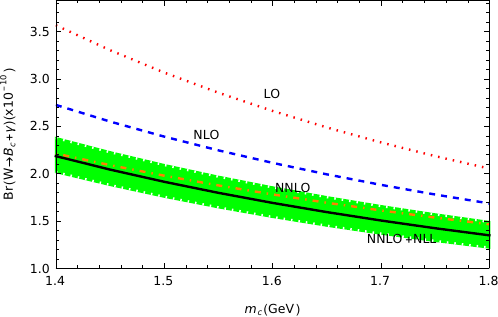}\hfill 
    \includegraphics[scale=0.88]{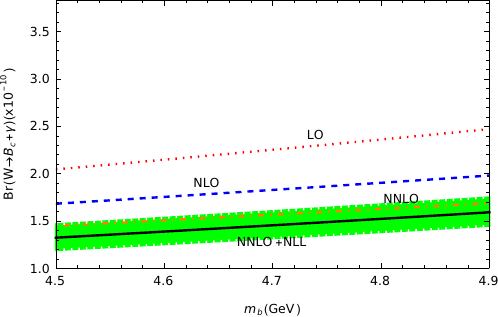}
    \caption{Variations of the branching fraction of $W\to B_c+\gamma$ with respect to the charm- and bottom-quark pole masses. The green band denotes the uncertainty from $\Gamma_W$, $|V_{cb}|$, and $\mu_R$. \label{fig-branching-ratio-mc-mb}}
\end{figure}

In figure~\ref{fig-branching-ratio-mc-mb}, we plot the branching fraction of $W\to B_c+\gamma$ as a function of the heavy-quark pole masses $m_c$ and $m_b$ at various levels of accuracy, where the green band denotes the uncertainty arising from the total decay width $\Gamma_W$, the CKM matrix element $|V_{cb}|$, and the renormalization scale $\mu_R$. We find that the branching fraction decreases monotonically as $m_c$ increases, while increasing monotonically as $m_b$ increases. Such a behavior is mainly attributed to the opposite dependence of the momentum fraction $x_0=\frac{m_c}{m_b+m_c}$ on the charm- and bottom-quark masses. Combing figure~\ref{fig-branching-ratio-mc-mb} with the data points collected in appendix~\ref{appendix-branching-ratios-mc-mb}, we can see that the branching ratios predicted at various orders are reduced by nearly a factor of two as the charm-quark mass $m_c$ increases. In contrast, the branching ratios have merely a small increase as the bottom-quark mass $m_b$ increases.

From the predicted decay widths and branching fractions at various levels of accuracy, we can see that the radiative decay $W\to B_c+\gamma$ is an extremely rare process. Although the LHC experiment has already accumulated a large amount of $W$-boson events, it is still impossible to observe this rare process in a foreseeable future. A huge number of $W$-boson events are also expected at the HL-LHC~\cite{Azzi:2019yne}, the FCC-ee~\cite{FCC:2018evy} and FCC-hh~\cite{FCC:2018vvp}, as well as the CEPC~\cite{CEPCStudyGroup:2018ghi}. For example, the FCC-hh is expected to produce $\mathcal{O}(10^{12})$ $W$ bosons~\cite{FCC:2018vvp,Mangano:2022ukr}, which would translate to hundreds of $W\to B_c+\gamma$ events according to our predictions. However, given the difficulty of reconstructing the $B_c$ signal, the experimental observation of this rare process is still very challenging. We look forward to further testing our theoretical predictions at these potential next-generation high-energy colliders. 

Finally, it should be noted that for a realistic measurement we need to consider both triggering and reconstruction. Since the photons produced in $W\to B_c+\gamma$ and $h\to\gamma\gamma$ processes have comparable energies, such events can definitely pass the triggering cuts on photon energy~\cite{Grossman:2015cak}. On the other hand, a straightforward reconstruction of the $B_c$ signal would be very difficult when considering the large amount of multijet background in a hadron collider. An alternative approach to study the rare $W\to B_c+\gamma$ decay is in a $t\bar{t}$ environment~\cite{Mangano:2014xta}. It offers a relatively cleaner environment with triggering on two $b$-jets and the leptonic decay of one $W$ boson, but the amount of $W$ bosons created through the $t\bar{t}$ production is much smaller. On the theoretical side, we should also note that for some ad-hoc experimental cuts that change the decay kinematics significantly, the power counting of the LC and NRQCD factorization approaches might be altered substantially. This would result in a conceptually quite different setup. We will be concerned with future experimental progress about these rare processes~\cite{CDF:1998kzn,LHCb:2022kta,CDF:2011jwr,CMS:2020oqe,ATLAS:2023jfq}.   

\section{Summary}
\label{sec-summary}

In this paper, we have performed a detailed theoretical study of the rare radiative decay $W\to B_c+\gamma$, by combining the NRQCD and LC factorization approaches. Especially, we have computed the NNLO QCD corrections to the NRQCD SDCs $\mathcal{C}_{1}$ and $\mathcal{C}_{2}$, and performed the LL and NLL resummations of large logarithms of $m_W^2/m_{B_c}^2$ present in these SDCs to all orders in $\alpha_s$. The decay width and branching fraction of $W\to B_c+\gamma$ are then analyzed phenomenologically, with all these corrections taken into account. It is found that, relative to the LO result, both the $\mathcal{O}(\alpha_s)$ and $\mathcal{O}(\alpha_s^2)$ corrections to the decay width are negative, with a net effect of $\sim-19\%$ and of $\sim-31\%$ respectively. Furthermore, while the LL resummation alters only slightly the NLO and NNLO predictions, the NLL resummation is always significant for the fixed-order NRQCD predictions at various orders in $\alpha_s$. We also found that the radiative corrections do not help to reduce the renormalization scale dependence of the branching fraction, which is however significantly reduced by the NLL resummation. Thus, we conclude that the NLL resummation is always important to improve the theoretical predictions of $W\to B_c+\gamma$ decay. Finally, we have investigated the variations of the decay width and branching fraction with respect to the heavy-quark pole masses $m_{b,c}$, find that the branching fraction decreases monotonically as $m_c$ increases, while increasing monotonically as $m_b$ increases. 

Although the radiative decay $W\to B_c+\gamma$ is an extremely rare process with a branching fraction of only $\mathcal{O}(10^{-10})$, we still expect that, with further advancements in experimental technologies, our theoretical predictions can be validated at the future next-generation high-energy colliders like FCC-hh.

\acknowledgments
This work is supported by the National Natural Science Foundation of China under Grant Nos.~12475094, 12135006 and 12075097, as well as by the Fundamental Research Funds for the Central Universities under Grant Nos.~CCNU22LJ004 and CCNU24AI003.


\appendix

\section{Explicit expressions of the ERBL kernels}
\label{appendix-BL-kernels}

The celebrated ERBL kernels at the one- and two-loop levels can be found, e.g., in refs.~\cite{Lepage:1979zb,Efremov:1978rn,Efremov:1979qk} and \cite{Sarmadi:1982yg,Dittes:1983dy,Katz:1984gf,Mikhailov:1984ii,Mueller:1993hg,Mueller:1994cn,Belitsky:1999gu}, respectively. For the non-singlet evolution up to $\mathcal{O}(\alpha_s^2)$, we have explicitly~\cite{Belitsky:1999gu}
\begin{eqnarray}
    V^{(0)}(x,y) &=& 2C_F\left[v(x,y)\right]_{+}\,,\\[0.2cm]
    V^{(1)}(x,y) &=& 4 C_F\Biggl[C_F V_F(x,y)-\frac{\beta_0}{2} V_\beta(x,y)-\left(C_F-\frac{C_A}{2}\right) V_G(x,y)\Biggr]_+ \nonumber \\[0.15cm]
    && -8C_F\Delta\beta_0 v^a(x,y)\,,
\end{eqnarray}
where
\begin{equation}
v(x,y) = f(x,y)\theta(y-x) + f(\bar{x},\bar{y})\theta(x-y)\,, \quad \text{with} \quad f(x,y) = \frac{x}{y}\left(1+\frac{1}{y-x}\right)\,,
\end{equation}
and
\begin{subequations}
	\begin{eqnarray}
	V_F(x, y) &= & \theta(y-x)\Biggl\{\left(\frac{4}{3}-2 \zeta(2)\right) f+3 \frac{x}{y}-\left(\frac{3}{2} f-\frac{x}{2 \bar{y}}\right) \ln 
        \frac{x}{y} - (f-\bar{f}) \ln \frac{x}{y} \ln \left(1-\frac{x}{y}\right) \nonumber \\
        && +\left(f+\frac{x}{2 \bar{y}}\right) \ln ^2 \frac{x}{y}\Biggr\} - \frac{x}{2 \bar{y}} \ln x(1+\ln x-2 \ln \bar{x}) + \left\{\begin{array}{l}
	x \rightarrow \bar{x} \\
	y \rightarrow \bar{y}
	\end{array}\right\}\,,\\[0.2cm]
	V_\beta(x,y) &= & \dot{v}(x,y)+\frac{5}{3}v(x,y)+v^a(x,y)\,,\\[0.2cm]
	V_G(x,y)&=& 2 v^a(x,y)+\frac{4}{3}v(x,y)\nonumber\\[0.1cm]
        &&+\Biggl[\theta(y-x)H(x,y)+\theta(x+y-1)\bar{H}(x,y) 
        + \left\{\begin{array}{l}
	x \rightarrow \bar{x} \\
	y \rightarrow \bar{y}
	\end{array}\right\}\Biggr]\,,
	\end{eqnarray}
\end{subequations}
with $f\equiv f(x,y)$, $\bar f\equiv  f(\bar x,\bar y)$, and
\begin{subequations}
	\begin{align}
	\dot{v}(x, y) &= \theta(y-x) f \ln \frac{x}{y}+\theta(x-y) \bar f\ln \frac{\bar x}{\bar y} \,,\\[0.2cm]
	v^a(x, y) &= \theta(y-x) \frac{x}{y}+\theta(x-y) \frac{\bar{x}}{\bar{y}}\,,\\[0.2cm]
	H(x, y) &= 2\Bigl[\bar{f}\left(\mathrm{Li}_2(\bar{x})+\ln y \ln \bar{x}\right)-f \operatorname{Li}_2(\bar{y})\Bigr]\,, \\[0.2cm]
	\bar{H}(x, y) &= 2\Biggl[(f-\bar{f})\left(\operatorname{Li}_2\left(1-\frac{x}{y}\right)+\frac{1}{2} \ln ^2 
        y\right)+f\left(\operatorname{Li}_2(\bar{y})-\operatorname{Li}_2(x)-\ln y \ln x\right)\Biggr]\,.
	\end{align}
\end{subequations}
In the above formulae, $\bar x = 1-x$ and $\bar y = 1-y$, $\zeta(s)=\sum_{n=1}^{\infty}1/n^s$ is the Riemann zeta function, and the familiar plus distribution is defined through
\begin{subequations}
	\begin{eqnarray}
	\int_0^1 dy \left[V(x,y)\right]_+ g(y) & = & \int_0^1 dy V(x,y)\left[g(y)-g(x)\right]\,,\\[0.3cm]
         \int_0^1 dx g(x) \left[V(x,y)\right]_+ & = & \int_0^1 dx \left[g(x)-g(y)\right] V(x,y)\,,
	\end{eqnarray}
\end{subequations}
for a test function $g(x)$.

\section{Analytical results of the convolutions}
\label{appendix-convolutions}

With the explicit expressions of the hard-scattering kernels $T_i(x,m_W,\mu)$, the leading-twist LCDA $\hat\phi(x,m_{B_c},\mu)$, and the ERBL evolution kernels $V^{(0,1)}(x,y)$, we can obtain the analytical results of the convolutions in eq.~\eqref{eq:NLL-expand}. Explicitly, we have  
\begin{subequations}
        \begin{eqnarray}
        \int_0^1 d x\, T_1^{(0)}(x) \hat\phi^{(0)}(x) &=& \frac{1}{4 \sqrt{2}}\Biggl\{\frac{e_c}{x_0}-\binom{e_c \rightarrow e_b}{x_0 \rightarrow \bar{x}_0}\Biggr\}-\frac{1}{2\sqrt{2}} \,,\\[0.2cm]
	\int_0^1 d x d y\, T_1^{(0)}(x) V^{(0)}(x, y) \hat\phi^{(0)}(y) &=& \frac{C_F}{4 \sqrt{2}} 
        \Biggl\{\frac{e_c(2\ln x_0+3)}{x_0}-\binom{e_c \rightarrow e_b}{x_0 \rightarrow \bar{x}_0}\Biggr\} \,,\nonumber \\ \\ 
	\int_0^1 d x d y d z\, T_1^{(0)}(x) V^{(0)}(x, y) V^{(0)}(y, z) \hat\phi^{(0)}(z) &=& \frac{C_F^2}{4 
        \sqrt{2}x_0\bar{x}_0}\Bigg\{e_c\Big[\bar{x}_0(4\ln^2 x_0-4\text{Li}_2(x_0)+9)\nonumber\\[0.1cm]
	&& \hspace{-1.8cm} -4(\bar{x}_0\ln \bar{x}_0+2x_0-3)\ln x_0\Big]-\binom{e_c \rightarrow e_b}{x_0 \rightarrow 
        \bar{x}_0}\Bigg\}\,,
	\end{eqnarray}
\end{subequations}
and 
\begin{subequations}
	\begin{eqnarray}
	\int_0^1 d x\, T_2^{(0)}(x) \hat\phi^{(0)}(x) &=& -\frac{1}{2 \sqrt{2}}\Biggl\{\frac{e_c}{x_0}+\binom{e_c \rightarrow 
        e_b}{x_0 \rightarrow \bar{x}_0}\Biggr\} \,, \\[0.2cm]
	\int_0^1 d x d y\, T_2^{(0)}(x) V^{(0)}(x, y) \hat\phi^{(0)}(y) &=& -\frac{C_F}{2 \sqrt{2}}
        \Biggl\{\frac{e_c(2\ln x_0+3)}{x_0}+\binom{e_c \rightarrow e_b}{x_0 \rightarrow \bar{x}_0}\Biggr\} \,,\nonumber\\ \\
	\int_0^1 d x d y d z\, T_2^{(0)}(x) V^{(0)}(x, y) V^{(0)}(y, z) \hat\phi^{(0)}(z) &=& \frac{C_F^2}{2 
        \sqrt{2}x_0\bar{x}_0}\Biggl\{e_c\Big[\bar{x}_0(4\text{Li}_2(x_0)-4\ln^2 x_0-9)\nonumber\\[0.1cm]
	&& \hspace{-1.8cm} +4(\bar{x}_0\ln \bar{x}_0+2x_0-3)\ln x_0\Big]+\binom{e_c \rightarrow e_b}{x_0 \rightarrow 
        \bar{x}_0}\Biggr\}\,,
	\end{eqnarray}
\end{subequations}
at the LL level,
\begin{subequations}
	\begin{eqnarray}
	\int_0^1 d x d y\, T_1^{(1)}(x) \hat\phi^{(0)}(y) &=& \frac{C_F}{4 \sqrt{2}x_0\bar{x}_0}\Bigg\{e_c\Big[\bar{x}_0\ln^2 x_0+\left((8\Delta-3)x_0-2i\pi\bar{x}_0\right)\ln x_0 \nonumber \\[0.1cm]
    && -(9+3i\pi)\bar{x}_0\Big] -\binom{e_c \rightarrow e_b}{x_0 \rightarrow \bar{x}_0}\Bigg\} 
    +\sqrt{2}C_F\Delta\,, \\[0.2cm]
	\int_0^1 d x d y\, T_1^{(0)}(x) \hat\phi^{(1)}(y) &=& \frac{C_F}{12 \sqrt{2}x_0\bar{x}_0}\Bigg\{e_c\Big[12\bar{x}_0\text{Li}_2(x_0)-6\bar{x}_0\ln^2 x_0-9\bar{x}_0\ln \bar{x}_0-15\bar{x}_0\ln x_0 \nonumber \\[0.1cm]
    && \hspace{-3.5cm} -24\Delta x_0\ln x_0-2\pi^2\bar{x}_0\Big] -\binom{e_c \rightarrow e_b}{x_0 \rightarrow \bar{x}_0}\Bigg\}+\frac{C_F}{2\sqrt{2}}\Bigg[(6x_0-3)\ln\Big(\frac{\bar{x}_0}{x_0}\Big)-4\Delta+6\Bigg]\,,
	\end{eqnarray}
\end{subequations}
and 
\begin{subequations}
	\begin{eqnarray}
	\int_0^1 d x d y\, T_2^{(1)}(x) \hat\phi^{(0)}(y) &=& -\frac{C_F}{2 \sqrt{2}x_0\bar{x}_0}\Bigg\{e_c\Big[\bar{x}_0\ln^2 x_0+((8\Delta-1)x_0-2i\pi\bar{x}_0)\ln x_0 \nonumber \\[0.1cm]
    && -(9+3i\pi)\bar{x}_0\Big] + \binom{e_c \rightarrow e_b}{x_0 \rightarrow \bar{x}_0}\Bigg\}\,, \\[0.2cm]
	\int_0^1 d x d y\, T_2^{(0)}(x) \hat\phi^{(1)}(y) &=& -\frac{C_F}{6 \sqrt{2}x_0\bar{x}_0}\Bigg\{e_c\Big[12\bar{x}_0\text{Li}_2(x_0)-6\bar{x}_0\ln^2 x_0-9\bar{x}_0\ln \bar{x}_0-15\bar{x}_0\ln x_0 \nonumber \\[0.1cm]
    && -24\Delta x_0\ln x_0-2\pi^2\bar{x}_0\Big] + \binom{e_c \rightarrow e_b}{x_0 \rightarrow \bar{x}_0}\Bigg\}\,,
	\end{eqnarray}
\end{subequations}
at the NLO level, as well as
\begin{subequations}
	\begin{eqnarray}
	&& \int_0^1 d x d y\, T_1^{(0)}(x) V^{(0)}(x, y) \hat\phi^{(1)}(y)=\frac{C_F^2}{12 \sqrt{2}x_0\bar{x}_0}\Bigg\{e_c\Big[2\pi^2 x_0-6\pi^2+(105x_0-24\Delta x_0\nonumber\\[0.1cm]
	&&~~
	-4\pi^2 \bar{x}_0-48 x_0^2-45)\ln x_0+(36x_0-24\Delta x_0-48)\ln^2 x_0-12\bar{x}_0\ln^3x_0+(48 x_0^2-45x_0\nonumber \\[0.2cm]
	&&~~
	-4\pi^2\bar{x}_0-6\bar{x}_0\ln x_0+24\bar{x}_0\text{Li}_2(x_0)-3)\ln\bar{x}_0+24\bar{x}_0\ln x_0\ln^2 \bar{x}_0+24(2-x_0)\text{Li}_2(x_0)\nonumber \\[0.1cm] 
        &&~~
    +24\bar{x}_0\text{Li}_3(\bar{x}_0)+48\bar{x}_0\text{Li}_3(x_0) -72\bar{x}_0\zeta(3)\Big]-\binom{e_c \rightarrow e_b}{x_0 \rightarrow \bar{x}_0}\Bigg\}\,,\\[0.2cm]
	&& \int_0^1 d x d y\, T_1^{(1)}(x) V^{(0)}(x, y) \hat\phi^{(0)}(y)=\frac{C_F^2}{4 \sqrt{2}x_0\bar{x}_0}\Bigg\{e_c\Big[-9(3+i\pi)\bar{x}_0+(-18-12i\pi+11 x_0\nonumber\\[0.1cm]
	&&~~
	+8(\Delta+i\pi) x_0)\ln x_0+(3-4i\pi \bar{x}_0-4(1-2\Delta)x_0)\ln^2 x_0+2\bar{x}_0\ln^3 x_0+(4i\pi \bar{x}_0\ln x_0\nonumber \\[0.1cm]
	&&~~
	-2\bar{x}_0\ln^2x_0)\ln \bar{x}_0+4i\pi \bar{x}_0\text{Li}_2(x_0)-4\bar{x}_0\text{Li}_3(x_0)+8\bar{x}_0\zeta(3)\Big]-\binom{e_c \rightarrow e_b}{x_0 \rightarrow \bar{x}_0}\Bigg\}\,, \\[0.2cm]
	&& \int_0^1 d x d y\, T_1^{(0)}(x) V^{(1)}(x, y) \hat\phi^{(0)}(y)=\frac{C_F}{24 \sqrt{2}x_0\bar{x}_0}\Bigg\{e_c\Big[3(3-2\pi^2)C_F\bar{x}_0+(3+2\pi^2)\beta_0 \bar{x}_0\nonumber\\[0.1cm]
	&&~~
	+6C_A\bar{x}_0-24C_F\bar{x}_0\ln x_0\ln^2\bar{x}_0+12(4C_F-\beta_0)\bar{x}_0\text{Li}_2(x_0)+\ln x_0(-24C_F x_0\nonumber \\[0.2cm]
	&&~~
	+(16-4\pi^2\bar{x}_0-40 x_0)C_A+(20-32x_0 +192\Delta x_0)\beta_0-24(C_A-2C_F)\bar{x}_0\text{Li}_2(x_0))\nonumber \\[0.2cm]
	&&~~
	+\ln \bar{x}_0(4\pi^2C_F\bar{x}_0-12(\beta_0-4C_F)\bar{x}_0\ln x_0-24C_F\bar{x}_0(\text{Li}_2(x_0)+\text{Li}_2(\bar{x}_0))\nonumber \\[0.1cm]
	&&~~
	+12C_F\bar{x}_0\ln^2 x_0)+24C_F\bar{x}_0\text{Li}_3(\bar{x}_0)+(72C_A-120C_F)\bar{x}_0\text{Li}_3(x_0)\nonumber\\[0.1cm]
	&&~~
    -(72C_A-96C_F)\bar{x}_0\zeta(3)\Big]-\binom{e_c \rightarrow e_b}{x_0 \rightarrow \bar{x}_0}\Bigg\}+4\sqrt{2}\Delta C_F\beta_0\,,
	\end{eqnarray}
\end{subequations}
and 
\begin{subequations}
	\begin{eqnarray}
	&& \int_0^1 d x d y\, T_2^{(0)}(x) V^{(0)}(x, y) \hat\phi^{(1)}(y)=-\frac{C_F^2}{6 \sqrt{2}x_0\bar{x}_0}\Bigg\{e_c\Big[2\pi^2 x_0-6\pi^2+ 
        (105x_0-24\Delta x_0\nonumber\\[0.1cm]
	&&~~
	-4\pi^2 \bar{x}_0-48 x_0^2-45)\ln x_0+(36x_0-24\Delta x_0-48)\ln^2 x_0-12\bar{x}_0\ln^3x_0+(48 x_0^2-45x_0\nonumber \\[0.2cm]
	&&~~
	-4\pi^2\bar{x}_0-6\bar{x}_0\ln x_0+24\bar{x}_0\text{Li}_2(x_0)-3)\ln\bar{x}_0+24\bar{x}_0\ln x_0\ln^2 \bar{x}_0+24(2-
        x_0)\text{Li}_2(x_0)\nonumber \\[0.1cm]
        &&~~
        +24\bar{x}_0\text{Li}_3(\bar{x}_0)+48\bar{x}_0\text{Li}_3(x_0) -72\bar{x}_0\zeta(3)\Big]+\binom{e_c \rightarrow e_b}{x_0 \rightarrow \bar{x}_0}\Bigg\}\,,\\[0.2cm]
	&& \int_0^1 d x d y\, T_2^{(1)}(x) V^{(0)}(x, y) \hat\phi^{(0)}(y)=-\frac{C_F^2}{2 \sqrt{2}x_0\bar{x}_0}
        \Bigg\{e_c\Big[-9(3+i\pi)\bar{x}_0+(-18-12i\pi+13 x_0\nonumber\\[0.1cm]
	&&~~
	+8(\Delta+i\pi) x_0)\ln x_0+(3-4i\pi \bar{x}_0-2(1-4\Delta)x_0)\ln^2 x_0+2\bar{x}_0\ln^3 x_0+(4i\pi \bar{x}_0\ln x_0\nonumber \\[0.1cm]
	&&~~
	-2\bar{x}_0\ln^2x_0)\ln \bar{x}_0+4i\pi \bar{x}_0\text{Li}_2(x_0)-4\bar{x}_0\text{Li}_3(x_0)+8\bar{x}_0\zeta(3)\Big]+\binom{e_c 
        \rightarrow e_b}{x_0 \rightarrow \bar{x}_0}\Bigg\}\,, \\[0.2cm]
	&& \int_0^1 d x d y\, T_2^{(0)}(x) V^{(1)}(x, y) \hat\phi^{(0)}(y)=-\frac{C_F}{12 \sqrt{2}x_0\bar{x}_0}\Bigg\{e_c\Big[3(3- 
        2\pi^2)C_F\bar{x}_0+(3+2\pi^2)\beta_0 \bar{x}_0\nonumber\\[0.1cm]
	&&~~
	+6C_A\bar{x}_0-24C_F\bar{x}_0\ln x_0\ln^2\bar{x}_0+12(4C_F-\beta_0)\bar{x}_0\text{Li}_2(x_0)+\ln x_0(-24C_F x_0 \nonumber \\[0.2cm]
	&&~~
	+(20-32x_0+192\Delta x_0)\beta_0+(16-4\pi^2\bar{x}_0-40 x_0)C_A-24(C_A-2C_F)\bar{x}_0\text{Li}_2(x_0))\nonumber \\[0.2cm]
	&&~~
	+\ln \bar{x}_0(4\pi^2C_F\bar{x}_0-12(\beta_0-4C_F)\bar{x}_0\ln x_0-24C_F\bar{x}_0(\text{Li}_2(x_0)+\text{Li}_2(\bar{x}_0))
        \nonumber \\[0.1cm]
	&&~~
	+12C_F\bar{x}_0\ln^2 x_0)+24C_F\bar{x}_0\text{Li}_3(\bar{x}_0)+(72C_A-120C_F)\bar{x}_0\text{Li}_3(x_0)\nonumber \\[0.1cm]
        &&~~
        -(72C_A-96C_F)\bar{x}_0\zeta(3)\Big]+\binom{e_c \rightarrow e_b}{x_0 \rightarrow \bar{x}_0}\Bigg\}\,,
	\end{eqnarray}
\end{subequations}
at the NLL level. Here $\Delta=0~(1)$ in the NDR (HV) scheme, and $\text{Li}_n(x)$ is the polylogarithm.

\section{Decay widths and branching ratios for different quark masses}
\label{appendix-branching-ratios-mc-mb}

\begin{table}[t]
    \renewcommand{\arraystretch}{1.3}
    \tabcolsep=0.39cm 
    \begin{center}	
    \begin{adjustbox}{width=0.98\textwidth,center}
    \begin{tabular}{|c|c|c|c|c|c|c|c|c|c|c|c|c|c|c|c|c|c|c|}
    \hline \hline
    $m_c$~[GeV] 
    & \multicolumn{2}{|c|}{1.40} & \multicolumn{2}{|c|}{1.45} & \multicolumn{2}{|c|}{1.50} & \multicolumn{2}{|c|}{1.55} 
    & \multicolumn{2}{|c|}{1.60} & \multicolumn{2}{|c|}{1.65} & \multicolumn{2}{|c|}{1.70} & \multicolumn{2}{|c|}{1.75} 
    & \multicolumn{2}{|c|}{1.80} \\
    \hline
    Accuracy level 
    &$\Gamma$&Br&$\Gamma$&Br&$\Gamma$&Br&$\Gamma$&Br&$\Gamma$&Br&$\Gamma$&Br&$\Gamma$&Br&$\Gamma$&Br&$\Gamma$&Br\\
    \hline
    LO&0.74&3.57&0.69&3.30&0.64&3.07&0.60&2.85&0.55&2.66&0.52&2.49&0.49&2.33&0.46&2.18&0.43&2.05 \\
    \hline
    NLO&0.56&2.72&0.53&2.55&0.50&2.39&0.47&2.25&0.44&2.11&0.42&1.99&0.39&1.88&0.37&1.79&0.35&1.68 \\	
    \hline
    NNLO&0.46&2.21&0.44&2.09&0.41&1.98&0.39&1.87&0.37&1.78&0.35&1.69&0.34&1.61&0.32&1.54&0.31&1.46 \\
    \hline		
    NNLO+NLL&0.46&2.18&0.43&2.04&0.40&1.91&0.37&1.80&0.35&1.69&0.33&1.59&0.31&1.50&0.30&1.42&0.28&1.35 \\
    \hline \hline
    \end{tabular}
    \end{adjustbox}
    \caption{Variations of the decay width (in unit of $\text{eV}$) and branching ratio (in unit of $10^{-10}$) of $W\to B_c+\gamma$ with respect to $m_c$, with all the other parameters fixed at their default values. \label{tab-mc-dependence}}
    \end{center}
\end{table}

\begin{table}[t]
    \renewcommand{\arraystretch}{1.3}
    \tabcolsep=0.39cm 
    \begin{center}	
    \begin{adjustbox}{width=0.98\textwidth,center}
    \begin{tabular}{|c|c|c|c|c|c|c|c|c|c|c|c|c|c|c|c|c|c|c|}
    \hline \hline
    $m_b$~[GeV]
    & \multicolumn{2}{|c|}{4.50} & \multicolumn{2}{|c|}{4.55}&\multicolumn{2}{|c|}{4.60}&\multicolumn{2}{|c|}{4.65}&\multicolumn{2}{|c|}{4.70}&\multicolumn{2}{|c|}{4.75}&\multicolumn{2}{|c|}{4.80}&\multicolumn{2}{|c|}{4.85}&\multicolumn{2}{|c|}{4.90}\\
    \hline 
    Accuracy level
    &$\Gamma$&Br&$\Gamma$&Br&$\Gamma$&Br&$\Gamma$&Br&$\Gamma$&Br&$\Gamma$&Br&$\Gamma$&Br&$\Gamma$&Br&$\Gamma$&Br\\
    \hline
    LO&0.43&2.04&0.44&2.10&0.45&2.15&0.46&2.20&0.47&2.25&0.48&2.31&0.49&2.36&0.50&2.41&0.51&2.47 \\
    \hline		
    NLO&0.35&1.68&0.36&1.72&0.37&1.75&0.37&1.79&0.38&1.83&0.39&1.86&0.40&1.90&0.40&1.94&0.41&1.98 \\
    \hline		
    NNLO&0.30&1.45&0.31&1.48&0.31&1.51&0.32&1.54&0.33&1.57&0.33&1.60&0.34&1.63&0.35&1.66&0.35&1.69 \\
    \hline		
    NNLO+NLL&0.28&1.32&0.28&1.36&0.29&1.39&0.30&1.42&0.30&1.45&0.31&1.49&0.32&1.52&0.32&1.56&0.33&1.59 \\
    \hline \hline
    \end{tabular}
    \end{adjustbox}
    \caption{Same as in table~\ref{tab-mc-dependence} but with respect to the bottom-quark pole mass $m_b$. 
    \label{tab-mb-dependence}}
    \end{center}
\end{table}

In order to analyze the variations of the observables of $W\to B_c+\gamma$ with respect to the heavy-quark pole masses $m_{b,c}$, we collect in tables~\ref{tab-mc-dependence} and \ref{tab-mb-dependence} the resulting decay widths and branching ratios by selecting eighteen distinct points of $m_c$ and $m_b$. 

\bibliographystyle{JHEP}
\bibliography{W2Bc_final-v3}

\providecommand{\href}[2]{#2}\begingroup\raggedright\begin{thebibliography}{100}

\bibitem{QuarkoniumWorkingGroup:2004kpm}
{\bf Quarkonium Working Group} Collaboration, N.~Brambilla et~al., {\it {Heavy quarkonium physics}},  \href{http://arxiv.org/abs/hep-ph/0412158}{{\tt hep-ph/0412158}}.

\bibitem{Brambilla:2010cs}
N.~Brambilla et~al., {\it {Heavy Quarkonium: Progress, Puzzles, and Opportunities}},  {\it Eur. Phys. J. C} {\bf 71} (2011) 1534, [\href{http://arxiv.org/abs/1010.5827}{{\tt arXiv:1010.5827}}].

\bibitem{Brambilla:2004jw}
N.~Brambilla, A.~Pineda, J.~Soto, and A.~Vairo, {\it {Effective Field Theories for Heavy Quarkonium}},  {\it Rev. Mod. Phys.} {\bf 77} (2005) 1423, [\href{http://arxiv.org/abs/hep-ph/0410047}{{\tt hep-ph/0410047}}].

\bibitem{CDF:1998ihx}
{\bf CDF} Collaboration, F.~Abe et~al., {\it {Observation of the $B_c$ meson in $p\bar{p}$ collisions at $\sqrt{s} = 1.8$ TeV}},  {\it Phys. Rev. Lett.} {\bf 81} (1998) 2432--2437, [\href{http://arxiv.org/abs/hep-ex/9805034}{{\tt hep-ex/9805034}}].

\bibitem{CDF:1998axz}
{\bf CDF} Collaboration, F.~Abe et~al., {\it {Observation of $B_c$ mesons in $p\bar{p}$ collisions at $\sqrt{s} = 1.8$ TeV}},  {\it Phys. Rev. D} {\bf 58} (1998) 112004, [\href{http://arxiv.org/abs/hep-ex/9804014}{{\tt hep-ex/9804014}}].

\bibitem{Chang:1992jb}
C.-H. Chang and Y.-Q. Chen, {\it {The hadronic production of the B(c) meson at Tevatron, CERN LHC and SSC}},  {\it Phys. Rev. D} {\bf 48} (1993) 4086--4091.

\bibitem{Berezhnoy:2019yei}
A.~V. Berezhnoy, I.~N. Belov, A.~K. Likhoded, and A.~V. Luhinsky, {\it {$B_c$ excitations at LHC experiments}},  {\it Mod. Phys. Lett. A} {\bf 34} (2019), no.~40 1950331, [\href{http://arxiv.org/abs/1904.06732}{{\tt arXiv:1904.06732}}].

\bibitem{Zheng:2015ixa}
X.-C. Zheng, C.-H. Chang, and Z.~Pan, {\it {Production of doubly heavy-flavored hadrons at $e^+e^-$ colliders}},  {\it Phys. Rev. D} {\bf 93} (2016), no.~3 034019, [\href{http://arxiv.org/abs/1510.06808}{{\tt arXiv:1510.06808}}].

\bibitem{Berezhnoy:2016etd}
A.~V. Berezhnoy, A.~K. Likhoded, A.~I. Onishchenko, and S.~V. Poslavsky, {\it {Next-to-leading order QCD corrections to paired $B_c$ production in $e^+e^-$ annihilation}},  {\it Nucl. Phys. B} {\bf 915} (2017) 224--242, [\href{http://arxiv.org/abs/1610.00354}{{\tt arXiv:1610.00354}}].

\bibitem{Zheng:2017xgj}
X.-C. Zheng, C.-H. Chang, T.-F. Feng, and Z.~Pan, {\it {NLO QCD corrections to $B_{c}$($B^\ast_{c}$) production around the $Z$ pole at an e$^{+}$ e$^{-}$ collider}},  {\it Sci. China Phys. Mech. Astron.} {\bf 61} (2018), no.~3 031012, [\href{http://arxiv.org/abs/1701.04561}{{\tt arXiv:1701.04561}}].

\bibitem{Zheng:2019gnb}
X.-C. Zheng, C.-H. Chang, T.-F. Feng, and X.-G. Wu, {\it {QCD NLO fragmentation functions for $c$ or $\bar{b}$ quark to $B_c$ or $B_c^*$ meson and their application}},  {\it Phys. Rev. D} {\bf 100} (2019), no.~3 034004, [\href{http://arxiv.org/abs/1901.03477}{{\tt arXiv:1901.03477}}].

\bibitem{Zhang:2021ypo}
Z.-Y. Zhang, X.-C. Zheng, and X.-G. Wu, {\it {Production of the $B_c$ meson at the CEPC}},  {\it Eur. Phys. J. C} {\bf 82} (2022), no.~3 246, [\href{http://arxiv.org/abs/2111.13917}{{\tt arXiv:2111.13917}}].

\bibitem{Zhan:2022etq}
X.-J. Zhan, X.-G. Wu, and X.-C. Zheng, {\it {Photoproduction of the Bc meson at future $e^+e^-$ colliders}},  {\it Phys. Rev. D} {\bf 106} (2022), no.~9 094036, [\href{http://arxiv.org/abs/2211.09003}{{\tt arXiv:2211.09003}}].

\bibitem{Bi:2016vbt}
H.-Y. Bi, R.-Y. Zhang, H.-Y. Han, Y.~Jiang, and X.-G. Wu, {\it {Photoproduction of the $B_c^{(*)}$ meson at the LHeC}},  {\it Phys. Rev. D} {\bf 95} (2017), no.~3 034019, [\href{http://arxiv.org/abs/1612.07990}{{\tt arXiv:1612.07990}}].

\bibitem{Chen:2020dtu}
Z.-Q. Chen, H.~Yang, and C.-F. Qiao, {\it {NLO QCD corrections to $B_c$-pair production in photon-photon collision}},  {\it Phys. Rev. D} {\bf 102} (2020), no.~1 016011, [\href{http://arxiv.org/abs/2005.07317}{{\tt arXiv:2005.07317}}].

\bibitem{Gouz:2002kk}
I.~P. Gouz, V.~V. Kiselev, A.~K. Likhoded, V.~I. Romanovsky, and O.~P. Yushchenko, {\it {Prospects for the $B_c$ studies at LHCb}},  {\it Phys. Atom. Nucl.} {\bf 67} (2004) 1559--1570, [\href{http://arxiv.org/abs/hep-ph/0211432}{{\tt hep-ph/0211432}}].

\bibitem{Gao:2010zzc}
Y.-N. Gao, J.~He, P.~Robbe, M.-H. Schune, and Z.-W. Yang, {\it {Experimental prospects of the $B_c$ studies of the LHCb experiment}},  {\it Chin. Phys. Lett.} {\bf 27} (2010) 061302.

\bibitem{HeavyFlavorAveragingGroupHFLAV:2024ctg}
{\bf Heavy Flavor Averaging Group (HFLAV)} Collaboration, S.~Banerjee et~al., {\it {Averages of $b$-hadron, $c$-hadron, and $\tau$-lepton properties as of 2023}},  \href{http://arxiv.org/abs/2411.18639}{{\tt arXiv:2411.18639}}.

\bibitem{ParticleDataGroup:2024cfk}
{\bf Particle Data Group} Collaboration, S.~Navas et~al., {\it {Review of particle physics}},  {\it Phys. Rev. D} {\bf 110} (2024), no.~3 030001.

\bibitem{UA1:1983crd}
{\bf UA1} Collaboration, G.~Arnison et~al., {\it {Experimental Observation of Isolated Large Transverse Energy Electrons with Associated Missing Energy at $\sqrt{s}= 540$ GeV}},  {\it Phys. Lett. B} {\bf 122} (1983) 103--116.

\bibitem{UA2:1983tsx}
{\bf UA2} Collaboration, M.~Banner et~al., {\it {Observation of Single Isolated Electrons of High Transverse Momentum in Events with Missing Transverse Energy at the CERN anti-p p Collider}},  {\it Phys. Lett. B} {\bf 122} (1983) 476--485.

\bibitem{Arnellos:1981gy}
L.~Arnellos, W.~J. Marciano, and Z.~Parsa, {\it {Radiative Decays $W^\pm \to \rho^\pm \gamma$ and $Z^0 \to \rho^0 \gamma$}},  {\it Nucl. Phys. B} {\bf 196} (1982) 378--393.

\bibitem{Keum:1993eb}
Y.~Y. Keum and X.-Y. Pham, {\it {Possible huge enhancement in the radiative decay of the weak $W$ boson into the charmed $D_s$ meson}},  {\it Mod. Phys. Lett. A} {\bf 9} (1994) 1545--1556, [\href{http://arxiv.org/abs/hep-ph/9303300}{{\tt hep-ph/9303300}}].

\bibitem{Grossman:2015cak}
Y.~Grossman, M.~K\"onig, and M.~Neubert, {\it {Exclusive Radiative Decays of W and Z Bosons in QCD Factorization}},  {\it JHEP} {\bf 04} (2015) 101, [\href{http://arxiv.org/abs/1501.06569}{{\tt arXiv:1501.06569}}].

\bibitem{Feng:2019meh}
F.~Feng, Y.~Jia, and W.-L. Sang, {\it {Optimized predictions for $W \to B_c + \gamma$ by combining light-cone and NRQCD approaches}},  \href{http://arxiv.org/abs/1902.11288}{{\tt arXiv:1902.11288}}.

\bibitem{Ishaq:2019zki}
S.~Ishaq, Y.~Jia, X.~Xiong, and D.-S. Yang, {\it {$W$ radiative decay to heavy-light mesons in HQET factorization through ${\cal O}(\alpha_s)$}},  {\it Phys. Rev. D} {\bf 100} (2019), no.~5 054027, [\href{http://arxiv.org/abs/1903.12627}{{\tt arXiv:1903.12627}}].

\bibitem{Bakos:2022lek}
E.~Bakos, N.~de~Groot, and N.~Vranjes, {\it {Identifying D Mesons from Radiative W Decays at the Large Hadron Collider}},  {\it Symmetry} {\bf 15} (2023), no.~10 1948, [\href{http://arxiv.org/abs/2207.13587}{{\tt arXiv:2207.13587}}].

\bibitem{Beneke:2023nmj}
M.~Beneke, G.~Finauri, K.~K. Vos, and Y.~Wei, {\it {QCD light-cone distribution amplitudes of heavy mesons from boosted HQET}},  {\it JHEP} {\bf 09} (2023) 066, [\href{http://arxiv.org/abs/2305.06401}{{\tt arXiv:2305.06401}}].

\bibitem{Bagdatova:2023etj}
A.~G. Bagdatova, S.~P. Baranov, and A.~S. Sakharov, {\it {Study of exclusive two-body W decays with fully reconstructible kinematics}},  {\it Mod. Phys. Lett. A} {\bf 38} (2023), no.~14n15 2350073, [\href{http://arxiv.org/abs/2303.02136}{{\tt arXiv:2303.02136}}].

\bibitem{Jones:2020bvu}
E.~Jones and W.~J. Murray, {\it {Mass biases in exclusive radiative hadronic decays of W bosons at the LHC}},  {\it New J. Phys.} {\bf 23} (2021), no.~11 113035, [\href{http://arxiv.org/abs/2009.01073}{{\tt arXiv:2009.01073}}].

\bibitem{Mangano:2014xta}
M.~Mangano and T.~Melia, {\it {Rare exclusive hadronic W decays in a $t\bar{t}$ environment}},  {\it Eur. Phys. J. C} {\bf 75} (2015), no.~6 258, [\href{http://arxiv.org/abs/1410.7475}{{\tt arXiv:1410.7475}}].

\bibitem{CDF:1998kzn}
{\bf CDF} Collaboration, F.~Abe et~al., {\it {Search for the rare decay $W^\pm \to D_s^{\pm} \gamma$ in $p\bar{p}$ collisions at $\sqrt{s} = 1.8$ TeV}},  {\it Phys. Rev. D} {\bf 58} (1998) 091101.

\bibitem{LHCb:2022kta}
{\bf LHCb} Collaboration, R.~Aaij et~al., {\it {Search for the rare decays $W^+ \to D^+_s\gamma$ and $Z \to D^0\gamma$ at LHCb}},  {\it Chin. Phys. C} {\bf 47} (2023), no.~9 093002, [\href{http://arxiv.org/abs/2212.07120}{{\tt arXiv:2212.07120}}].

\bibitem{CDF:2011jwr}
{\bf CDF} Collaboration, T.~Aaltonen et~al., {\it {Search for the Rare Radiative Decay: $W\rightarrow\pi\gamma$ in $p\bar{p}$ Collisions at $\sqrt{s} = 1.96$ TeV}},  {\it Phys. Rev. D} {\bf 85} (2012) 032001, [\href{http://arxiv.org/abs/1104.1585}{{\tt arXiv:1104.1585}}].

\bibitem{CMS:2020oqe}
{\bf CMS} Collaboration, A.~M. Sirunyan et~al., {\it {Search for the rare decay of the W boson into a pion and a photon in proton-proton collisions at $\sqrt{s}=13$~TeV}},  {\it Phys. Lett. B} {\bf 819} (2021) 136409, [\href{http://arxiv.org/abs/2011.06028}{{\tt arXiv:2011.06028}}].

\bibitem{ATLAS:2023jfq}
{\bf ATLAS} Collaboration, G.~Aad et~al., {\it {Search for the Exclusive $W$ Boson Hadronic Decays $W^{\pm}\to \pi^{\pm}\gamma$, $W^{\pm}\to K^{\pm}\gamma$ and $W^{\pm}\to \rho^{\pm}\gamma$ with the ATLAS Detector}},  {\it Phys. Rev. Lett.} {\bf 133} (2024), no.~16 161804, [\href{http://arxiv.org/abs/2309.15887}{{\tt arXiv:2309.15887}}].

\bibitem{Azzi:2019yne}
P.~Azzi et~al., {\it {Report from Working Group 1}: {Standard Model Physics at the HL-LHC and HE-LHC}},  {\it CERN Yellow Rep. Monogr.} {\bf 7} (2019) 1--220, [\href{http://arxiv.org/abs/1902.04070}{{\tt arXiv:1902.04070}}].

\bibitem{FCC:2018evy}
{\bf FCC} Collaboration, A.~Abada et~al., {\it {FCC-ee: The Lepton Collider}: {Future Circular Collider Conceptual Design Report Volume 2}},  {\it Eur. Phys. J. ST} {\bf 228} (2019), no.~2 261--623.

\bibitem{FCC:2018vvp}
{\bf FCC} Collaboration, A.~Abada et~al., {\it {FCC-hh: The Hadron Collider}: {Future Circular Collider Conceptual Design Report Volume 3}},  {\it Eur. Phys. J. ST} {\bf 228} (2019), no.~4 755--1107.

\bibitem{CEPCStudyGroup:2018ghi}
{\bf CEPC Study Group} Collaboration, M.~Dong et~al., {\it {CEPC Conceptual Design Report: Volume 2 - Physics \& Detector}},  \href{http://arxiv.org/abs/1811.10545}{{\tt arXiv:1811.10545}}.

\bibitem{dEnterria:2023wjq}
D.~d'Enterria and V.~D. Le, {\it {Rare and exclusive few-body decays of the Higgs, Z, W bosons, and the top quark}},  \href{http://arxiv.org/abs/2312.11211}{{\tt arXiv:2312.11211}}.

\bibitem{Lepage:1980fj}
G.~P. Lepage and S.~J. Brodsky, {\it {Exclusive Processes in Perturbative Quantum Chromodynamics}},  {\it Phys. Rev. D} {\bf 22} (1980) 2157.

\bibitem{Chernyak:1983ej}
V.~L. Chernyak and A.~R. Zhitnitsky, {\it {Asymptotic Behavior of Exclusive Processes in QCD}},  {\it Phys. Rept.} {\bf 112} (1984) 173.

\bibitem{Lepage:1979zb}
G.~P. Lepage and S.~J. Brodsky, {\it {Exclusive Processes in Quantum Chromodynamics: Evolution Equations for Hadronic Wave Functions and the Form-Factors of Mesons}},  {\it Phys. Lett. B} {\bf 87} (1979) 359--365.

\bibitem{Efremov:1978rn}
A.~V. Efremov and A.~V. Radyushkin, {\it {Asymptotical Behavior of Pion Electromagnetic Form-Factor in QCD}},  {\it Theor. Math. Phys.} {\bf 42} (1980) 97--110.

\bibitem{Efremov:1979qk}
A.~V. Efremov and A.~V. Radyushkin, {\it {Factorization and Asymptotical Behavior of Pion Form-Factor in QCD}},  {\it Phys. Lett. B} {\bf 94} (1980) 245--250.

\bibitem{Guberina:1980dc}
B.~Guberina, J.~H. Kuhn, R.~D. Peccei, and R.~Ruckl, {\it {Rare Decays of the $Z^0$}},  {\it Nucl. Phys. B} {\bf 174} (1980) 317--334.

\bibitem{Wang:2013ywc}
X.-P. Wang and D.~Yang, {\it {The leading twist light-cone distribution amplitudes for the S-wave and P-wave quarkonia and their applications in single quarkonium exclusive productions}},  {\it JHEP} {\bf 06} (2014) 121, [\href{http://arxiv.org/abs/1401.0122}{{\tt arXiv:1401.0122}}].

\bibitem{Huang:2014cxa}
T.-C. Huang and F.~Petriello, {\it {Rare exclusive decays of the Z-boson revisited}},  {\it Phys. Rev. D} {\bf 92} (2015), no.~1 014007, [\href{http://arxiv.org/abs/1411.5924}{{\tt arXiv:1411.5924}}].

\bibitem{Bodwin:2017pzj}
G.~T. Bodwin, H.~S. Chung, J.-H. Ee, and J.~Lee, {\it {$Z$-boson decays to a vector quarkonium plus a photon}},  {\it Phys. Rev. D} {\bf 97} (2018), no.~1 016009, [\href{http://arxiv.org/abs/1709.09320}{{\tt arXiv:1709.09320}}].

\bibitem{Chung:2019ota}
H.~S. Chung, J.-H. Ee, D.~Kang, U.-R. Kim, J.~Lee, and X.-P. Wang, {\it {Pseudoscalar Quarkonium+gamma Production at NLL+NLO accuracy}},  {\it JHEP} {\bf 10} (2019) 162, [\href{http://arxiv.org/abs/1906.03275}{{\tt arXiv:1906.03275}}].

\bibitem{Sang:2022erv}
W.-L. Sang, D.-S. Yang, and Y.-D. Zhang, {\it {Z boson radiative decays to a P-wave quarkonium at NNLO and LL accuracy}},  {\it Phys. Rev. D} {\bf 106} (2022), no.~9 094023, [\href{http://arxiv.org/abs/2208.10118}{{\tt arXiv:2208.10118}}].

\bibitem{Sang:2023hjl}
W.-L. Sang, D.-S. Yang, and Y.-D. Zhang, {\it {Z-boson radiative decays to an S-wave quarkonium at NNLO and NLL accuracy}},  {\it Phys. Rev. D} {\bf 108} (2023), no.~1 014021, [\href{http://arxiv.org/abs/2302.06439}{{\tt arXiv:2302.06439}}].

\bibitem{Bodwin:1994jh}
G.~T. Bodwin, E.~Braaten, and G.~P. Lepage, {\it {Rigorous QCD analysis of inclusive annihilation and production of heavy quarkonium}},  {\it Phys. Rev. D} {\bf 51} (1995) 1125--1171, [\href{http://arxiv.org/abs/hep-ph/9407339}{{\tt hep-ph/9407339}}]. [Erratum: Phys.Rev.D 55, 5853 (1997)].

\bibitem{Chen:2015csa}
L.-B. Chen and C.-F. Qiao, {\it {Two-loop QCD Corrections to $B_c$ Meson Leptonic Decays}},  {\it Phys. Lett. B} {\bf 748} (2015) 443--450, [\href{http://arxiv.org/abs/1503.05122}{{\tt arXiv:1503.05122}}].

\bibitem{Tao:2023mtw}
W.~Tao, Z.-J. Xiao, and R.~Zhu, {\it {Three-loop matching coefficients for heavy flavor-changing currents and the phenomenological applications}},  {\it JHEP} {\bf 05} (2023) 189, [\href{http://arxiv.org/abs/2303.07220}{{\tt arXiv:2303.07220}}].

\bibitem{Bodwin:2002cfe}
G.~T. Bodwin and A.~Petrelli, {\it {Order-$v^4$ corrections to $S$-wave quarkonium decay}},  {\it Phys. Rev. D} {\bf 66} (2002) 094011, [\href{http://arxiv.org/abs/hep-ph/0205210}{{\tt hep-ph/0205210}}]. [Erratum: Phys.Rev.D 87, 039902 (2013)].

\bibitem{Kublbeck:1990xc}
J.~Kublbeck, M.~Bohm, and A.~Denner, {\it {Feyn Arts: Computer Algebraic Generation of Feynman Graphs and Amplitudes}},  {\it Comput. Phys. Commun.} {\bf 60} (1990) 165--180.

\bibitem{Hahn:2000kx}
T.~Hahn, {\it {Generating Feynman diagrams and amplitudes with FeynArts 3}},  {\it Comput. Phys. Commun.} {\bf 140} (2001) 418--431, [\href{http://arxiv.org/abs/hep-ph/0012260}{{\tt hep-ph/0012260}}].

\bibitem{Mertig:1990an}
R.~Mertig, M.~Bohm, and A.~Denner, {\it {FEYN CALC: Computer algebraic calculation of Feynman amplitudes}},  {\it Comput. Phys. Commun.} {\bf 64} (1991) 345--359.

\bibitem{Shtabovenko:2016sxi}
V.~Shtabovenko, R.~Mertig, and F.~Orellana, {\it {New Developments in FeynCalc 9.0}},  {\it Comput. Phys. Commun.} {\bf 207} (2016) 432--444, [\href{http://arxiv.org/abs/1601.01167}{{\tt arXiv:1601.01167}}].

\bibitem{Shtabovenko:2020gxv}
V.~Shtabovenko, R.~Mertig, and F.~Orellana, {\it {FeynCalc 9.3: New features and improvements}},  {\it Comput. Phys. Commun.} {\bf 256} (2020) 107478, [\href{http://arxiv.org/abs/2001.04407}{{\tt arXiv:2001.04407}}].

\bibitem{Shtabovenko:2023idz}
V.~Shtabovenko, R.~Mertig, and F.~Orellana, {\it {FeynCalc 10: Do multiloop integrals dream of computer codes?}},  {\it Comput. Phys. Commun.} {\bf 306} (2025) 109357, [\href{http://arxiv.org/abs/2312.14089}{{\tt arXiv:2312.14089}}].

\bibitem{Feng:2012tk}
F.~Feng and R.~Mertig, {\it {FormLink/FeynCalcFormLink: Embedding FORM in Mathematica and FeynCalc}},  \href{http://arxiv.org/abs/1212.3522}{{\tt arXiv:1212.3522}}.

\bibitem{Beneke:1997zp}
M.~Beneke and V.~A. Smirnov, {\it {Asymptotic expansion of Feynman integrals near threshold}},  {\it Nucl. Phys. B} {\bf 522} (1998) 321--344, [\href{http://arxiv.org/abs/hep-ph/9711391}{{\tt hep-ph/9711391}}].

\bibitem{Smirnov:2002pj}
V.~A. Smirnov, {\it {Applied asymptotic expansions in momenta and masses}},  {\it Springer Tracts Mod. Phys.} {\bf 177} (2002) 1--262.

\bibitem{Beneke:1997jm}
M.~Beneke, A.~Signer, and V.~A. Smirnov, {\it {Two loop correction to the leptonic decay of quarkonium}},  {\it Phys. Rev. Lett.} {\bf 80} (1998) 2535--2538, [\href{http://arxiv.org/abs/hep-ph/9712302}{{\tt hep-ph/9712302}}].

\bibitem{Broadhurst:1991fy}
D.~J. Broadhurst, N.~Gray, and K.~Schilcher, {\it {Gauge invariant on-shell Z(2) in QED, QCD and the effective field theory of a static quark}},  {\it Z. Phys. C} {\bf 52} (1991) 111--122.

\bibitem{Melnikov:2000zc}
K.~Melnikov and T.~van Ritbergen, {\it {The Three loop on-shell renormalization of QCD and QED}},  {\it Nucl. Phys. B} {\bf 591} (2000) 515--546, [\href{http://arxiv.org/abs/hep-ph/0005131}{{\tt hep-ph/0005131}}].

\bibitem{Fael:2020bgs}
M.~Fael, K.~Sch\"onwald, and M.~Steinhauser, {\it {Exact results for $ {Z}_m^{\mathrm{OS}} $ and $ {Z}_2^{\mathrm{OS}} $ with two mass scales and up to three loops}},  {\it JHEP} {\bf 10} (2020) 087, [\href{http://arxiv.org/abs/2008.01102}{{\tt arXiv:2008.01102}}].

\bibitem{Kreimer:1989ke}
D.~Kreimer, {\it {The $\gamma_5$ Problem and Anomalies: A Clifford Algebra Approach}},  {\it Phys. Lett. B} {\bf 237} (1990) 59--62.

\bibitem{Korner:1991sx}
J.~G. Korner, D.~Kreimer, and K.~Schilcher, {\it {A Practicable $\gamma_5$ scheme in dimensional regularization}},  {\it Z. Phys. C} {\bf 54} (1992) 503--512.

\bibitem{Feng:2012iq}
F.~Feng, {\it {$\tt{Apart}$: A Generalized Mathematica Apart Function}},  {\it Comput. Phys. Commun.} {\bf 183} (2012) 2158--2164, [\href{http://arxiv.org/abs/1204.2314}{{\tt arXiv:1204.2314}}].

\bibitem{Smirnov:2008iw}
A.~V. Smirnov, {\it {Algorithm FIRE -- Feynman Integral REduction}},  {\it JHEP} {\bf 10} (2008) 107, [\href{http://arxiv.org/abs/0807.3243}{{\tt arXiv:0807.3243}}].

\bibitem{Smirnov:2014hma}
A.~V. Smirnov, {\it {FIRE5: A C++ implementation of Feynman Integral REduction}},  {\it Comput. Phys. Commun.} {\bf 189} (2015) 182--191, [\href{http://arxiv.org/abs/1408.2372}{{\tt arXiv:1408.2372}}].

\bibitem{Smirnov:2019qkx}
A.~V. Smirnov and F.~S. Chukharev, {\it {FIRE6: Feynman Integral REduction with modular arithmetic}},  {\it Comput. Phys. Commun.} {\bf 247} (2020) 106877, [\href{http://arxiv.org/abs/1901.07808}{{\tt arXiv:1901.07808}}].

\bibitem{Smirnov:2023yhb}
A.~V. Smirnov and M.~Zeng, {\it {FIRE 6.5: Feynman integral reduction with new simplification library}},  {\it Comput. Phys. Commun.} {\bf 302} (2024) 109261, [\href{http://arxiv.org/abs/2311.02370}{{\tt arXiv:2311.02370}}].

\bibitem{Patel:2015tea}
H.~H. Patel, {\it {Package-X: A Mathematica package for the analytic calculation of one-loop integrals}},  {\it Comput. Phys. Commun.} {\bf 197} (2015) 276--290, [\href{http://arxiv.org/abs/1503.01469}{{\tt arXiv:1503.01469}}].

\bibitem{Patel:2016fam}
H.~H. Patel, {\it {Package-X 2.0: A Mathematica package for the analytic calculation of one-loop integrals}},  {\it Comput. Phys. Commun.} {\bf 218} (2017) 66--70, [\href{http://arxiv.org/abs/1612.00009}{{\tt arXiv:1612.00009}}].

\bibitem{Liu:2017jxz}
X.~Liu, Y.-Q. Ma, and C.-Y. Wang, {\it {A Systematic and Efficient Method to Compute Multi-loop Master Integrals}},  {\it Phys. Lett. B} {\bf 779} (2018) 353--357, [\href{http://arxiv.org/abs/1711.09572}{{\tt arXiv:1711.09572}}].

\bibitem{Liu:2020kpc}
X.~Liu, Y.-Q. Ma, W.~Tao, and P.~Zhang, {\it {Calculation of Feynman loop integration and phase-space integration via auxiliary mass flow}},  {\it Chin. Phys. C} {\bf 45} (2021), no.~1 013115, [\href{http://arxiv.org/abs/2009.07987}{{\tt arXiv:2009.07987}}].

\bibitem{Liu:2021wks}
X.~Liu and Y.-Q. Ma, {\it {Multiloop corrections for collider processes using auxiliary mass flow}},  {\it Phys. Rev. D} {\bf 105} (2022), no.~5 L051503, [\href{http://arxiv.org/abs/2107.01864}{{\tt arXiv:2107.01864}}].

\bibitem{Davydychev:1992mt}
A.~I. Davydychev and J.~B. Tausk, {\it {Two loop selfenergy diagrams with different masses and the momentum expansion}},  {\it Nucl. Phys. B} {\bf 397} (1993) 123--142.

\bibitem{Broadhurst:1993mw}
D.~J. Broadhurst, J.~Fleischer, and O.~V. Tarasov, {\it {Two loop two point functions with masses: Asymptotic expansions and Taylor series, in any dimension}},  {\it Z. Phys. C} {\bf 60} (1993) 287--302, [\href{http://arxiv.org/abs/hep-ph/9304303}{{\tt hep-ph/9304303}}].

\bibitem{Tarasov:1997kx}
O.~V. Tarasov, {\it {Generalized recurrence relations for two loop propagator integrals with arbitrary masses}},  {\it Nucl. Phys. B} {\bf 502} (1997) 455--482, [\href{http://arxiv.org/abs/hep-ph/9703319}{{\tt hep-ph/9703319}}].

\bibitem{Martin:2003qz}
S.~P. Martin, {\it {Evaluation of two loop selfenergy basis integrals using differential equations}},  {\it Phys. Rev. D} {\bf 68} (2003) 075002, [\href{http://arxiv.org/abs/hep-ph/0307101}{{\tt hep-ph/0307101}}].

\bibitem{Schroder:2005va}
Y.~Schroder and A.~Vuorinen, {\it {High-precision epsilon expansions of single-mass-scale four-loop vacuum bubbles}},  {\it JHEP} {\bf 06} (2005) 051, [\href{http://arxiv.org/abs/hep-ph/0503209}{{\tt hep-ph/0503209}}].

\bibitem{Baikov:2010hf}
P.~A. Baikov and K.~G. Chetyrkin, {\it {Four Loop Massless Propagators: An Algebraic Evaluation of All Master Integrals}},  {\it Nucl. Phys. B} {\bf 837} (2010) 186--220, [\href{http://arxiv.org/abs/1004.1153}{{\tt arXiv:1004.1153}}].

\bibitem{Lee:2011jt}
R.~N. Lee, A.~V. Smirnov, and V.~A. Smirnov, {\it {Master Integrals for Four-Loop Massless Propagators up to Transcendentality Weight Twelve}},  {\it Nucl. Phys. B} {\bf 856} (2012) 95--110, [\href{http://arxiv.org/abs/1108.0732}{{\tt arXiv:1108.0732}}].

\bibitem{Liu:2022chg}
X.~Liu and Y.-Q. Ma, {\it {AMFlow: A Mathematica package for Feynman integrals computation via auxiliary mass flow}},  {\it Comput. Phys. Commun.} {\bf 283} (2023) 108565, [\href{http://arxiv.org/abs/2201.11669}{{\tt arXiv:2201.11669}}].

\bibitem{Liu:2022mfb}
Z.-F. Liu and Y.-Q. Ma, {\it {Determining Feynman Integrals with Only Input from Linear Algebra}},  {\it Phys. Rev. Lett.} {\bf 129} (2022), no.~22 222001, [\href{http://arxiv.org/abs/2201.11637}{{\tt arXiv:2201.11637}}].

\bibitem{Chetyrkin:2000yt}
K.~G. Chetyrkin, J.~H. Kuhn, and M.~Steinhauser, {\it {RunDec: A Mathematica package for running and decoupling of the strong coupling and quark masses}},  {\it Comput. Phys. Commun.} {\bf 133} (2000) 43--65, [\href{http://arxiv.org/abs/hep-ph/0004189}{{\tt hep-ph/0004189}}].

\bibitem{Herren:2017osy}
F.~Herren and M.~Steinhauser, {\it {Version 3 of RunDec and CRunDec}},  {\it Comput. Phys. Commun.} {\bf 224} (2018) 333--345, [\href{http://arxiv.org/abs/1703.03751}{{\tt arXiv:1703.03751}}].

\bibitem{Ma:2006hc}
J.~P. Ma and Z.~G. Si, {\it {NRQCD Factorization for Twist-2 Light-Cone Wave-Functions of Charmonia}},  {\it Phys. Lett. B} {\bf 647} (2007) 419--426, [\href{http://arxiv.org/abs/hep-ph/0608221}{{\tt hep-ph/0608221}}].

\bibitem{Bell:2008er}
G.~Bell and T.~Feldmann, {\it {Modelling light-cone distribution amplitudes from non-relativistic bound states}},  {\it JHEP} {\bf 04} (2008) 061, [\href{http://arxiv.org/abs/0802.2221}{{\tt arXiv:0802.2221}}].

\bibitem{Jia:2008ep}
Y.~Jia and D.~Yang, {\it {Refactorizing NRQCD short-distance coefficients in exclusive quarkonium production}},  {\it Nucl. Phys. B} {\bf 814} (2009) 217--230, [\href{http://arxiv.org/abs/0812.1965}{{\tt arXiv:0812.1965}}].

\bibitem{Jia:2010fw}
Y.~Jia, J.-X. Wang, and D.~Yang, {\it {Bridging light-cone and NRQCD approaches: asymptotic behavior of $B_c$ electromagnetic form factor}},  {\it JHEP} {\bf 10} (2011) 105, [\href{http://arxiv.org/abs/1012.6007}{{\tt arXiv:1012.6007}}].

\bibitem{Xu:2016dgp}
J.~Xu and D.~Yang, {\it {The leading twist light-cone distribution amplitudes for the S-wave and P-wave B$_{c}$ mesons}},  {\it JHEP} {\bf 07} (2016) 098, [\href{http://arxiv.org/abs/1604.04441}{{\tt arXiv:1604.04441}}].

\bibitem{tHooft:1972tcz}
G.~'t~Hooft and M.~J.~G. Veltman, {\it {Regularization and Renormalization of Gauge Fields}},  {\it Nucl. Phys. B} {\bf 44} (1972) 189--213.

\bibitem{Agaev:2010aq}
S.~S. Agaev, V.~M. Braun, N.~Offen, and F.~A. Porkert, {\it {Light Cone Sum Rules for the pi0-gamma*-gamma Form Factor Revisited}},  {\it Phys. Rev. D} {\bf 83} (2011) 054020, [\href{http://arxiv.org/abs/1012.4671}{{\tt arXiv:1012.4671}}].

\bibitem{Bodwin:2016edd}
G.~T. Bodwin, H.~S. Chung, J.-H. Ee, and J.~Lee, {\it {New approach to the resummation of logarithms in Higgs-boson decays to a vector quarkonium plus a photon}},  {\it Phys. Rev. D} {\bf 95} (2017), no.~5 054018, [\href{http://arxiv.org/abs/1603.06793}{{\tt arXiv:1603.06793}}].

\bibitem{Eichten:1995ch}
E.~J. Eichten and C.~Quigg, {\it {Quarkonium wave functions at the origin}},  {\it Phys. Rev. D} {\bf 52} (1995) 1726--1728, [\href{http://arxiv.org/abs/hep-ph/9503356}{{\tt hep-ph/9503356}}].

\bibitem{Mangano:2022ukr}
M.~Aleksa et~al., {\it {Conceptual design of an experiment at the FCC-hh, a future 100 TeV hadron collider}}, .

\bibitem{Sarmadi:1982yg}
M.~H. Sarmadi, {\it {The Asymptotic Pion Form-factor Beyond the Leading Order}},  {\it Phys. Lett. B} {\bf 143} (1984) 471.

\bibitem{Dittes:1983dy}
F.~M. Dittes and A.~V. Radyushkin, {\it {TWO LOOP CONTRIBUTION TO THE EVOLUTION OF THE PION WAVE FUNCTION}},  {\it Phys. Lett. B} {\bf 134} (1984) 359--362.

\bibitem{Katz:1984gf}
G.~R. Katz, {\it {Two Loop Feynman Gauge Calculation of the Meson Nonsinglet Evolution Potential}},  {\it Phys. Rev. D} {\bf 31} (1985) 652.

\bibitem{Mikhailov:1984ii}
S.~V. Mikhailov and A.~V. Radyushkin, {\it {Evolution Kernels in {QCD}: Two Loop Calculation in Feynman Gauge}},  {\it Nucl. Phys. B} {\bf 254} (1985) 89--126.

\bibitem{Mueller:1993hg}
D.~Mueller, {\it {Conformal constraints and the evolution of the nonsinglet meson distribution amplitude}},  {\it Phys. Rev. D} {\bf 49} (1994) 2525--2535.

\bibitem{Mueller:1994cn}
D.~Mueller, {\it {The Evolution of the pion distribution amplitude in next-to-leading-order}},  {\it Phys. Rev. D} {\bf 51} (1995) 3855--3864, [\href{http://arxiv.org/abs/hep-ph/9411338}{{\tt hep-ph/9411338}}].

\bibitem{Belitsky:1999gu}
A.~V. Belitsky, D.~Mueller, and A.~Freund, {\it {Reconstruction of nonforward evolution kernels}},  {\it Phys. Lett. B} {\bf 461} (1999) 270--279, [\href{http://arxiv.org/abs/hep-ph/9904477}{{\tt hep-ph/9904477}}].

\end{thebibliography}\endgroup

\end{document}